\documentstyle[psfig,mncite]{mn}

\newif\ifAMStwofonts

\newcommand{\ea}{E$+$A}

\newcommand{\hb}{H$\beta$}

\newcommand{\hg}{H$\gamma$}
\newcommand{\hga}{H$\gamma_A$}
\newcommand{\hst}{{\it HST}}
\newcommand{\kms}{km\,s$^{-1}$}
\newcommand{\mgb}{Mg$_b$}

\ifoldfss
  \ifCUPmtlplainloaded \else
    \NewTextAlphabet{textbfit} {cmbxti10} {}
    \NewTextAlphabet{textbfss} {cmssbx10} {}
    \NewMathAlphabet{mathbfit} {cmbxti10} {} 
    \NewMathAlphabet{mathbfss} {cmssbx10} {} 
  \fi
  \ifAMStwofonts
    \ifCUPmtlplainloaded \else
      \NewSymbolFont{upmath} {eurm10}
      \NewSymbolFont{AMSa} {msam10}
      \NewMathSymbol{\upi}     {0}{upmath}{19}
      \NewMathSymbol{\umu}     {0}{upmath}{16}
      \NewMathSymbol{\upartial}{0}{upmath}{40}
      \NewMathSymbol{\leqslant}{3}{AMSa}{36}
      \NewMathSymbol{\geqslant}{3}{AMSa}{3E}

      \let\geq=\geqslant 
    \fi
  \fi
\fi 

\ifnfssone
  \newmathalphabet{\mathit}
  \addtoversion{normal}{\mathit}{cmr}{m}{it}
  \addtoversion{bold}{\mathit}{cmr}{bx}{it}
  \newmathalphabet{\mathbfit} 
  \addtoversion{normal}{\mathbfit}{cmr}{bx}{it}
  \addtoversion{bold}{\mathbfit}{cmr}{bx}{it}
  \newmathalphabet{\mathbfss} 
  \addtoversion{normal}{\mathbfss}{cmss}{bx}{n}
  \addtoversion{bold}{\mathbfss}{cmss}{bx}{n}
  \ifAMStwofonts
    \ifCUPmtlplainloaded \else
      \UseAMStwoboldmath
      \makeatletter
      \new@mathgroup\upmath@group
      \define@mathgroup\mv@normal\upmath@group{eur}{m}{n}
      \define@mathgroup\mv@bold\upmath@group{eur}{b}{n}
      \edef\UPM{\hexnumber\upmath@group}
      \new@mathgroup\amsa@group
      \define@mathgroup\mv@normal\amsa@group{msa}{m}{n}
      \define@mathgroup\mv@bold\amsa@group{msa}{m}{n}
      \edef\AMSa{\hexnumber\amsa@group}
      \makeatother
      \mathchardef\upi="0\UPM19
      \mathchardef\umu="0\UPM16
      \mathchardef\upartial="0\UPM40
      \mathchardef\leqslant="3\AMSa36
      \mathchardef\geqslant="3\AMSa3E

      \let\geq=\geqslant 
    \fi
  \fi
\fi 

\ifnfsstwo
  \DeclareMathAlphabet{\mathbfit}{OT1}{cmr}{bx}{it}
  \SetMathAlphabet\mathbfit{bold}{OT1}{cmr}{bx}{it}
  \DeclareMathAlphabet{\mathbfss}{OT1}{cmss}{bx}{n}
  \SetMathAlphabet\mathbfss{bold}{OT1}{cmss}{bx}{n}
  \ifAMStwofonts
    \ifCUPmtlplainloaded \else
      \DeclareSymbolFont{UPM}{U}{eur}{m}{n}
      \SetSymbolFont{UPM}{bold}{U}{eur}{b}{n}
      \DeclareSymbolFont{AMSa}{U}{msa}{m}{n}
      \DeclareMathSymbol{\upi}{0}{UPM}{"19}
      \DeclareMathSymbol{\umu}{0}{UPM}{"16}
      \DeclareMathSymbol{\upartial}{0}{UPM}{"40}
      \DeclareMathSymbol{\leqslant}{3}{AMSa}{"36}
      \DeclareMathSymbol{\geqslant}{3}{AMSa}{"3E}

      \let\geq=\geqslant 
    \fi
  \fi
\fi 

\ifCUPmtlplainloaded \else
  \ifAMStwofonts \else 
    \def\upi{\pi}
    \def\umu{\mu}
    \def\upartial{\partial}
  \fi
\fi

\title{The early--type galaxy population in Abell\,2218}
\author[B. Ziegler et al.]
       {Bodo L.\ Ziegler\thanks{Present address: Universit\"atssternwarte,
Geismarlandstr. 11, 37083 G\"ottingen, Germany, 
E-mail: bziegler@uni-sw.gwdg.de}, 
Richard G.\ Bower, Ian Smail, Roger
L.\ Davies \& David Lee\thanks{Present address: Anglo-Australian Observatory, Epping, Sydney, NSW, Australia} \\
        Department of Physics, University of Durham, Durham DH1\,3LE, UK}
\date{Accepted 2000 .
      Received 2000 ;
      in original form 2000 }

\pubyear{2000}

\begin{document}
\bibliographystyle{abbrv}

\maketitle

\label{firstpage}

\begin{abstract}
We present high signal-to-noise, moderate-resolution spectroscopy of 48
early-type members of the rich cluster Abell\,2218 at $z=0.18$ taken
with the LDSS2 spectrograph on the 4.2-m William Herschel Telescope.
This sample is both larger and spans a wider galaxy luminosity range,
down to $M^\ast_B+1$, than previous studies.  In addition to the
relatively large size of the sample we have detailed morphological
information from archival {\it Hubble Space Telescope} imaging for 20
of the galaxies.  We combine the morphological, photometric, kinematic
and line-strength information to compare A\,2218 with similar samples drawn from local clusters and to identify
evolutionary changes between the samples which have occured over the
last $\approx3$\,Gyrs.  The overall picture is one of little or no
evolution in nearly all galaxy parameters. Zeropoint offsets in the
Faber--Jackson, \mgb--$\sigma$ and Fundamental Plane relations are all
consistent with passively evolving stellar populations. The slopes of
these relations have not changed significantly  in the 3\,Gyrs
between A\,2218 and today.  We do however find a significant spread in
the estimated luminosity-weighted ages of the stellar populations in
the galaxies, based on line diagnostic diagrams. This age spread is
seen in both the disky early-type galaxies (S0) and also  the
ellipticals.  We observe both ellipticals with a strong contribution
from a young stellar population and lenticulars dominated by old
stellar populations.  On average, we find no evidence for systematic
differences between the populations of ellipticals and lenticulars.  
In both
cases there appears to be little evidence for differences between the
stellar populations of the two samples.  This points  to a common
formation epoch for the bulk of the stars in most of the early--type
galaxies in A\,2218. This result can be reconciled with the claims
of rapid morphological evolution in distant clusters 
if the suggested transformation from spirals to lenticulars does not
involve significant new star formation. 
\end{abstract}


\section{Introduction}

The application of local galaxy scaling relations to the galaxy
populations in distant clusters is widely regarded as one of the most
powerful routes to understanding their formation and development.  The
main scaling relations which have been employed are the Faber--Jackson
relation (FJR, luminosity--velocity dispersion, \pcite{FJ76}), the
Kormendy relation (KR, surface brightness--effective radius,
\pcite{Korme77}), the Fundamental Plane (FP, combining surface
brightness, effective radius and velocity dispersion,
\pcite{DLBDFTW87,DD87,BBF92}) and the Mg--velocity
dispersion relation (Mg--$\sigma$, \pcite{BBF93,CBDMSW99}).  These four
relationships test different aspects of the luminosity and mass
evolution of cluster galaxies and are predominantly applicable to the
early-type galaxies which dominate the cores of rich clusters both
locally and at high redshifts.
 
Two of these relations rely on accurate photometric and structural
information for the galaxies, and as result there has been significant
work employing WFPC2 on-board {\it Hubble Space Telescope} ({\it
HST}\,) to study the photometric evolution of
morphologically-classified elliptical galaxies in distant clusters
using the KR (\pcite{BSL96}; Schade et al.\ 1996, 1997;
\pcite{FCAF98,BASECDOPS98,ZSBBGS99}).  These studies suggest mild
evolution in luminosity and size for ellipticals in rich clusters,
consistent with passive evolution.
Other groups assume ``a priori'' a degree of luminosity evolution based
upon stellar population models and then attempt to use the observed KR
evolution to constrain cosmological models \cite{PDC96,MCKFB98}.
Unfortunately, as was pointed out by \scite{ZSBBGS99}, the statistical
and systematic errors in the measurement and transformation of  {\it
HST} magnitudes are too large to give unambigous results for
cosmological parameters.

More detailed information comes from studies investigating the
evolution of the FJR and FP which supplement the photometry (both {\it
HST} and ground-based) with moderate-resolution spectroscopy, capable
of resolving velocity dispersions down to $\approx100$\,\kms.   Results
have so far been published on the  clusters A\,2218 ($z=0.17$) and
A\,665 ($z=0.18$) by \scite{JFHD99}; MS\,1358$+$62 ($z=0.33$) and
MS\,2053--04 ($z=0.58$) by \scite{KDFIF97}; A\,370, MS\,1512$+$36 and
Cl\,0949$+$44 all at $z=0.37$ (see \pcite{BZB96,ZB97,BSZBBGH97b});
Cl\,0024$+$16 ($z=0.39$) by \scite{DF96}, and MS\,1054--03 ($z=0.83$)
by \scite{DFKI98}.  In particular, the study of
\scite{KIDF00} exploited a mosaic of {\it HST} images and  Keck
spectroscopy, to construct the FP for 30 luminous E and S0 galaxies in
Cl\,1358$+$62. The shape of the FP is very similar to that of the local
Coma cluster and the offset between them can be explained by a mild
evolution in luminosity. Unfortunately line strengths could not be
explored since the widely-used Mg$_b$ index falls into a strong
telluric band at the cluster redshift.  The combined data from all
these studies strongly favour  passive evolution of the stellar
populations of early-type cluster galaxies both in surface brightness
and mass-to-light ratio and a high redshift of formation, $z>2$.
Further support for a high formation epoch for the stellar populations
in early-type cluster galaxies comes from the modest evolution in the
\mgb\ absorption at a constant velocity dispersion in galaxies out to
$z=0.37$ \cite{ZB97}.

The relatively modest evolution claimed for the early-type population
in clusters is in strong contrast to the  evolution seen in the
morphological mix in these environments.  In particular, the origin of
the well-studied ``Butcher-Oemler'' effect \cite{BO84} in a population
of star-forming cluster galaxies with spiral morphologies is now well
established \cite{CESS94,CBSES98}.  More recently there have been claims
of morphological evolution in some classes of early-type galaxies in
rich clusters.  {\it HST} images have revealed an increasing paucity
of S0 (lenticular) galaxies with redshift \cite{DOCSE97}. While S0s
form the dominant luminous galaxy population in local rich clusters,
\scite{DOCSE97} found  that only 10--20\% of bright cluster galaxies
are S0s in rich clusters at $z\approx0.5$.  They suggested that
this strong increase to the present-day resulted from the gradual
transformation of spiral galaxies,  accreted from the surrounding
field, into S0s \cite{PSDCB99}. If this transformation is relatively
rapid, $<2$--3\,Gyrs, then some proportion of the S0 population may
show signatures of this past star formation activity in terms of
young stellar populations and blue colours.  However, low resolution
spectroscopic surveys have typically failed to find large populations
of bright, blue S0s \cite{CBSES98}.  Moreover, the S0 population as a whole 
ought to show a wider dispersion in age than the cluster ellipticals.
However, photometric analysis of morphologically-classified samples of
ellipticals and S0s in distant clusters shows them to be almost identical
in their 4000\AA\ break colours \cite{E97}, and hence by
implication their ages.  Unfortunately, the degeneracy between
age and metallicity in most observables based on integrated spectra
(e.g.\ broad-band optical photometry) make these hypotheses difficult
to test conclusively \cite{Worth94}.

To break the age--metallicity degeneracy a number of groups have tried to
identify absorption lines which are predominantly sensitive to age, e.g.\
Balmer lines, or which depend strongly on metallicity, mostly combinations
of metal lines, \cite{JW95,CVPB96,WO97,VA99}.  Good progress has been
made and we can now  construct line diagnostic diagrams in which the
age--metallicity degeneracy is nearly broken \cite{KD98,Joerg99}.  Recent
analysis based on such line indices for the composite S0 populations in
distant clusters have indicated that the stellar populations in luminous
lenticular and elliptical cluster galaxies at $z=0.31$ have very similar
luminosity-weighted mean ages \cite{JSC00}.  This suggests that  a model
with a rapid transformation, from star-forming spiral to passive S0,
is unlikely to be correct and a more gradual mechanism may be required
\cite{PSDCB99,KS00}.

Another approach to breaking the age--metallicity degeneracy is to use
a combination of optical and optical-infrared photometry \cite{A78}.
This approach has been used very recently by \scite{S00} to study the
luminosity-weighted ages of early-type galaxies spanning a very wide
range in luminosity in A\,2218.  This analysis suggests
that while the most luminous ellipticals and S0s have very similar
luminosity-weighted ages, the less luminous S0s may exhibit a much
wider range in ages.  Likewise, an intensive spectroscopic analysis of
individual galaxies in the local Fornax poor cluster by \scite{Kunt00}
has found that some of the lower luminosity S0 galaxies have extended star
formation histories, compared to the more luminous ellipticals and S0s.

In the light of the discussion above, it is clearly important to apply
the detailed spectroscopic techniques to individual galaxies spanning
as wide a range in luminosity as possible in distant clusters, to study
their line strengths and kinematics.  However, 
this is very difficult and most previous spectroscopic studies
have targetted only a modest number of typically the more luminous (and
hence massive) galaxies in each cluster.  We have therefore undertaken
a programme to obtain high-quality spectra of a large number of
early-type galaxies (of order 50) across a wide range of luminosity in
two rich clusters A\,2218 and A\,2390.  Both clusters have been imaged
by {\it HST} providing accurate structural parameters.  At $z=0.18$ and
$z=0.23$ respectively, the cluster galaxies are  bright enough to
observe even sub-$L^*$ systems with 4-m class telescopes, while still
representing a look-back time of around a quarter of the age of the
Universe, adequate to address evolutionary questions. Both clusters are
very rich systems and may well serve in the future as more suitable
benchmarks for the comparison to very rich, high redshift clusters
($z\approx1$) than Coma.  Moreover, comparisons between A\,2218 and
A\,2390 and more distant systems also have the advantage that aperture
corrections are less important than for comparisons based on Coma.

In this paper, we present the study
of galaxies in A\,2218. In \S2, the observations and the data reduction
are described. Galaxy scaling relations and line diagnostic diagrams
are examined in \S3, and in \S4 we discuss our results and give our
conclusions in \S5. The paper ends with  comprehensive Appendices listing
all our observational data.

Throughout the paper we adopt the following values for the cosmological
parameters: $H_0=60$\,km\,s$^{-1}$\,Mpc$^{-1}$, $q_0=0.1$,
$\Lambda=0$.  For the nearby Coma comparison cluster ($z=0.024$) this
results in a distance modulus of $dm=35.42$\,mag and for A\,2218
($z=0.175$) in $dm=40.0$\,mag, a scale of 3.51\,kpc\,arcsec$^{-1}$ and
a look--back time of 2.6\,Gyrs.

\section{Observations and data reduction}

A\,2218 is  a very rich cluster at $z=0.18$ with a large velocity
dispersion,  $\sigma=1370^{+160}_{-120}$\,\kms \cite{LPS92}
and a high X-ray luminosity of $L_X$(0.5--4.4\,keV) = $6.5\times
10^{44}$ ergs\,s$^{-1}$.  An {\it HST}-based lensing analysis provides
a detailed view of the mass distribution within the central 1\,Mpc of
the cluster and indicates a mass of $M(r<400{\rm kpc}) = 4 \times
10^{14} M_\odot$ \cite{KESCS96}.  Most interestingly, the lensing
map identifies two mass concentrations within the core of the cluster
and led to the suggestion that it is suffering (or has just suffered) a
core-penetrating merger.

A\,2218 has previously been observed in a Fundamental Plane study by
J{\o}rgensen et al.\ (1999, JFHD).  They targetted a sample of 11 early-type
galaxies photometrically-selected from relatively shallow ground-based
imaging. Their main conclusion was that both the evolution in luminosity, as 
well as in mass-to-light ratio, is modest and in agreement with passive 
stellar population models (e.g.\ \pcite{BC93}).

Our study of the early-type galaxy population in A\,2218 is based upon
optical photometry from the 5.1-m Hale telescope at Palomar Observatory
and multi-object spectroscopy using LDSS2 at the 4.2-m William Herschel
Telescope (WHT) on La Palma. In addition, we have exploited WFPC2 images
taken with {\it HST}  to provide high-quality morphological information
for a subset of our sample.  We next discuss the properties of these
various datasets.

\subsection{Hale/COSMIC imaging}

The ground-based $UBVI$ imaging of a $9.7'\times 9.7'$ region centered
on A\,2218 (Fig.\,\ref{finder}) 
used for our initial target selection was obtained with
the COSMIC imaging spectrograph on the 5.1-m Hale telescope at Palomar
during June 1994 and June 1995. A thick 2048$^2$ TEK CCD (pixel scale
$0.286''$\,pixel$^{-1}$) was used and individual exposures of $\sim
500$--1000\,s were taken using in--field dithering (on a grid with $\sim
30''$ spacing). The frames were reduced in a standard manner with {\sc
iraf}\footnote{IRAF is distributed by the National Optical
Astronomy Observatories, which are operated by the Association of
Universities for Research in Astronomy, Inc., under cooperative
agreement with the National Science Foundation.}
using twilight flatfields to initially flatfield the frames before
masking the brighter objects and stacking the data frames to create
master flatfields in each filter.  The data were then flatfielded using
these master frames, aligned and coadded using a cosmic ray rejection
algorithm.   The total exposure times are 2.5\,ks in $U$, 16.5\,ks in $B$,
5.8\,ks in $V$ and 21.7\,ks in $I$ and the seeing measured on the final
frames was 1.20$''$, 1.25$''$, 2.05$''$ and 0.95$''$ FWHM for $UBVI$
respectively.  The frames were calibrated onto the Johnson/Cousins
system of \scite{Lo92} and corrected for reddening calculated from the
{\it COBE} dust maps \cite{SFD98}. We estimate $E(B-V)=0.025$ leading to
extinction coefficients for the Landolt filters of $A_U = 0.136, A_B =
0.108, A_V = 0.083, A_R = 0.067, A_I = 0.048$ and $A_{702} = 0.061$
for the {\it HST}/WFPC2 F702W filter.  We determine $1\sigma$ limiting
surface brightnesses of  $\mu_U = 26.1$, $\mu_B = 28.9$, $\mu_V = 28.0$
and $\mu_I = 27.4$ mag arcsec$^{-2}$ for the final ground-based frames.

We used the SExtractor image analysis package \cite{BA96} to create a
catalogue of galaxies detected in our deep $I$-band frame.  We adopted
a minimum area after convolution with a $0.86''$ FWHM top-hat filter of
0.82\,sq.\ arcsec above the $\mu_I = 24.6$ mag arcsec$^{-2}$ isophote.
The galaxy catalogue created in this way includes positions, total
magnitudes ({\sc mag\_best} from SExtractor) and  crude morphological
information on $\sim 2800$ stars and galaxies within the $I$-band
frame. To calculate colours for these galaxies we measured aperture
magnitudes within 4.0-$''$ (14\,kpc) diameter apertures on the
seeing-matched $UBVI$ frames and used these to construct $(U-B)$,
$(B-V)$ and $(B-I)$ colours for the galaxies. In the Appendix we list
the observed colours and coordinates for the galaxies we obtained
spectra for (Table\,\ref{coomag}). The galaxy identifications are
plotted on the full field in Fig.\,\ref{finder} and thumbnail images of
the 48 galaxies are given in Fig.\,\ref{mosaicp200}.

For the Faber--Jackson relation we
translate our observed $I$-band magnitudes into absolute Gunn $r$
magnitudes, $r_{\rm abs}$, following the description of JFHD. 
The transformation of
$I$ into restframe $r_{\rm rest}$ can be calculated with:
\begin{equation}
\label{rrest}
r_{\rm rest}=I+0.062 (V-I)+0.76
\end{equation}
Subtracting the distance modulus for A\,2218 from $r_{\rm rest}$ then yields 
$r_{\rm abs}$.
We use the extinction-corrected total $I$-band magnitudes ($I_{tot}$)
and aperture $(V-I)$ colours in this calculation.

\subsection{HST/WFPC2 imaging}

In addition to the deep multi-colour ground-based imaging,
morphological information is also available for a subset of the
galaxies in our sample.  This comes from an {\it HST} image of A\,2218
taken  with the WFPC2 camera on September 2, 1994.  The total exposure
time was 6.5\,ks in the F702W filter ($R_{702}$) and the final frame covers a
field of roughly $2.5'\times 2.5'$ at $\sim 0.15''$ resolution in the
core of the cluster. These data were reduced and analysed
by \scite{KESCS96} and more details can be found in that work. 

The WFPC2 image has a $1\sigma$ surface brightness limit of $\mu_{R}
\approx 30$ mag pixel$^{-2}$, which is more than adequate to provide
high-quality morphological information on the brighter cluster members.
Thumbnail images of the 19 galaxies (disregarding the cD) for which we
have spectra are shown in Fig.\,\ref{mosaichst} and we indicate the
position of the {\it HST} field on Fig.\,\ref{finder}.  The brighter
galaxies in the {\it HST} field ($R_{702}<22.5$) have been visually classified
by Prof.\ W.\,Couch onto the revised Hubble scheme used by the MORPHS
project (see \scite{S00}).

\subsubsection{Determination of structural parameters}
\label{dethst}

To measure the structural parameters of the
spec\-tro\-scopic\-ally-observed early-type galaxies lying within the
{\it HST} image (plus the cD galaxy) we extracted subframes of all 19
galaxies using {\sc midas}\footnote{ESO--MIDAS, the European Southern 
Observatory Munich Image Data Analysis System is developed and maintained by 
the European Southern Observatory.}
and analysed each galaxy using the profile
fitting method developed by Saglia et al. (1997a, 1997b).
We briefly summarise this technique here:
after masking stars and artifacts from around the
galaxy, the circularly
averaged surface brightness profile of the galaxy was fitted with
PSF-convolved $r^{1/4}$ and exponential components (both simultaneously
and separately).  The PSF was generated using the {\sc tinytim} software
package \cite{KH97}.  The quality of the fits was assessed by Monte
Carlo simulations, taking into account sky-subtraction corrections,
signal-to-noise, the radial extent of the profiles and the $\chi^2$
quality of the fit. In this way, we were able to derive the effective
radius ($r_e$ in arcsec), the total $F702W$-band magnitude ($R_{702}$)
and the mean surface brightness within $r_e$ ($\langle\mu_e\rangle$)
for the entire galaxy as well as the luminosity and scale of the bulge
($m_b$ and $r_{e,b}$) and disk ($m_d$ and $h$) component separately,
within the limitations described by \scite{SBBBCDMW97}. We estimate
the average error in $R_{702}$ and $r_e$ to be $0.015$\,mag and $25\%$ respectively.
The structural parameters for all the galaxies can be found in
Table\,\ref{hstpar} of the Appendix.

We also performed an isophote analysis based on the procedure
introduced by  \scite{BM87}. Deviations from the elliptical isophotes
were recorded as a function of radius by a Fourier decomposition
algorithm.  The presence and strength of the a4 coefficient (taken as a
signature of diskyness) is in good agreement with the visually
determined morphological class of S0 and early-type spirals.

\subsubsection{Rest-Frame Surface Brightness Profiles}
\label{rfsbp}

In order to determine the rest frame properties of the galaxies,
we perform the transformation of the {\it HST}
magnitudes to restframe Gunn $r_{rest}$ as described by JFHD. We summarize here the
transformations introduced by JFHD: According to \scite{HBCHTWW95}, the
instrumental magnitude in the F702W filter is:
\begin{equation}
R_{702} = -2.5 \log({\rm DN})/t_{{\rm exp}}+ZP+2.5\log(GR)
\end{equation}
with $GR$ being the respective gain ratios for the WF chips and $ZP$ the
zeropoint for an exposure time $t_{{\rm exp}}=1$\,s. $ZP=21.670$ with
consideration of the difference between ``short'' and ``long'' exposures of 
0.05\,mag \cite{H98} and an aperture correction of 0.109 \cite{HBCHTWW95}. This is transformed into Cousins $I$ using the equations
\begin{equation}
R_c = R_{702}+0.486 (V-R_c)-0.079 (V-R_c)^2
\end{equation}
and
\begin{equation}
(V-R_c)=0.52*(V-I)
\end{equation}
which yields:
\begin{equation}
I = R_{702} - 0.227 (V-I) - 0.021 (V-I)^2
\end{equation}
The $(V-I)$ colour was taken from our ground-based aperture photometry,
see Table\,\ref{coomag}. The $4''$ photometric aperture yields
an equivalent area for the distant galaxies to the area used in the
Coma photometry.  Finally, the calibration to restframe Gunn $r_{rest}$ is
achieved using Equation~1.

Since the observed F702W passband is close to restframe Gunn $r_{rest}$ at
the redshift of A\,2218, the overall k-corrections are small. On average
$R_{702}-r_{rest}=-0.40$.  For the same reason, the effective radii can be
directly compared to the ones measured of Coma galaxies in Gunn $r$. The
mean surface brightness within $r_e$ is:
\begin{equation}
\langle\mu_r\rangle_e = r_{rest}+2.5 \log (2\pi)+5 \log(r_e) - 10 \log(1+z)
\end{equation}
where the last term corrects for the dimming due to the expansion of the
Universe. The mean surface brightness in units of L$_o$/pc$^2$ is
\begin{equation}
\log \langle I\rangle_e = -0.4 (\langle\mu_r\rangle_e - 26.4)
\end{equation}
With an angular distance of A\,2218 of $d=683$\,Mpc for our cosmology, the 
effective radius in kpc is:
\begin{equation}
\log(R_e) = \log(r_e)+\log(d)-2.314 \quad (R_e \,\mbox{in kpc})
\end{equation}
whereas Coma ($z=0.024$) has $d=115.7$\,Mpc.

Compared to the data given by JFHD, our measurements are almost
identical. For this comparison, we used only their {\it HST}-based
sample. There are 15 galaxies in common with respect to the structural
parameters and the largest deviations in $\langle\mu_r\rangle_e$
for these are only 3--6\% in four galaxies (the cD and \#1454, \#1552
and \#1662). These deviations may be due to residual background light
from nearby bright galaxies (\#1552 and \#1662 are close to the cD,
\#1454 close to the giant elliptical \#1437). For the remainder of the
galaxies we summarize the comparison in Table\,\ref{cfhst}. This table
also contains the estimated total (random and systematic) errors of
the parameters.  While there are 15 galaxies between our sample and
that from JFHD available to compare structural parameters, there are
only six for which a similar spectroscopic comparison is possible. 

%
\begin{table}
\centering
\begin{minipage}{80mm}
\caption{Difference between this paper and JFHD.}
\label{cfhst}
\begin{tabular}{rrcll}
parameter & N$_{\mathrm{gal}}$ & Difference & error & error$_{\rm JFHD}$ \\
\hline
$R_{\mathrm{702}}$ & 11 & $-$0.01$\pm$0.12 & 0.015 & 0.015 \\
$(V-I)$ & 15 & 0.023$\pm$0.047 & 0.007 & 0.025 \\
$r_z$ & 11 & $-$0.03$\pm$0.11 & 0.05 & 0.05 \\
$\log r_e$ & 11 & 0.003$\pm$0.071 & 0.111 & 0.078 \\
$\langle\mu_r\rangle_e$ & 11 & $-$0.00$\pm$0.24 & 0.25 & 0.29 \\
$\log \sigma$ & 8 & 0.06$\pm$0.07 & 0.020 & 0.025 \\
\end{tabular}

\medskip
Errors are estimated total errors (random $+$ systematic).
\end{minipage}
\end{table}

\subsection{WHT/LDSS2 multi--object spectroscopy}

\subsubsection{Sample Selection}

The aim of the project is to study the stellar populations of a
large sample of early-type cluster galaxies spanning a wide range in
luminosity. However, owing to the need for good sky subtraction in
our spectra we were constrained to include only around 20 galaxies in
each mask.  For this reason we took great care to select only galaxies
which were  likely to be cluster members based upon their broad-band
colours from our ground-based $UBVI$ imaging.

\begin{figure}
\psfig{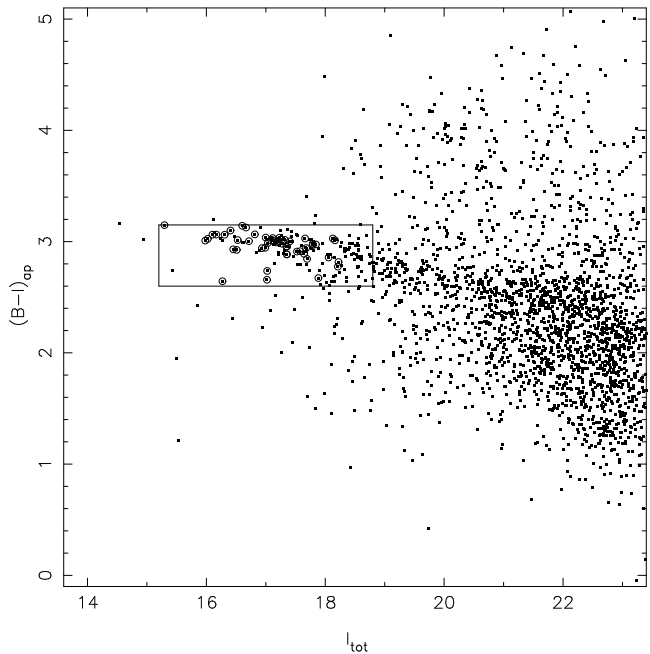}
\caption{The $(B-I)$--$I$ colour magnitude diagram from our Hale
imaging for bright galaxies lying in a $9.7'\times9.7'$
(2.2\,Mpc) region centered on A\,2218. The sequence of red cluster
members is readily seen extending down to $I\sim 22$ ($M_B\sim -17$).
The box encompasses galaxies matching our selection criteria, while
circles denote those galaxies which were actually observed.}
\label{ibi}
\end{figure}

Using the existing redshift catalogue for the field \cite{LPS92}, we
defined a region in $UBVI$ colour space which was occupied by cluster
members. Galaxies falling outside the colour region,
defined as $2.60<(B-I)<3.20$,  $-0.50<(B-V)<0.50$ and $-0.10<(U-B)<0.50$,
were rejected (see, e.g.\ Fig.\,\ref{ibi}).  The width of this region  places negligible restrictions on
the stellar populations of the selected galaxies, but rejects the
majority of background galaxies. Combined with the richness of the
cluster this ensured a high rate of success in targetting cluster
members in our spectroscopic sample (1 non-member out of 49 targets).
The  colour-selected sample of galaxies includes 138 galaxies with $I_{\rm
tot} < 18.8$ (our adopted magnitude limit) within the $9.7'\times9.7'$
area covered by the Hale images.

This catalogue was then used as the input for mask design.
We prepared two masks containing galaxies with an average $I$-band
surface brightness within the spectroscopic slit brighter than
$\mu_I=19.7$\,mag\,arcsec$^{-2}$. A third mask contained galaxies as faint
as  $\mu_I=20.2$\,mag\,arcsec$^{-2}$. Dividing the galaxies between the 
masks in this way
enabled exposure times to be chosen
to ensure that similar signal to noise was obtained in each spectrum.

The position angles of the masks were chosen so as to maximise the number
of galaxies with early-type morphologies  (E--S0--Sa) selected within the
{\it HST} field. Galaxies were allocated to slits manually.  With few
exceptions, a minimum slit length of $15''$ was used, and the position
of the slit was also chosen so that fainter companion galaxies did not
obscure the sky. Occasionally, the choice of either of two galaxies
made the packing of the mask equally efficient. In this situation,
we preferred galaxies for which a redshift was already known.

\subsubsection{Observations}
\label{obs}

During the nights of June 2--5, 1997, we obtained multi-object
spectroscopy using LDSS2 \cite{JRAS94} at the
WHT.  To obtain high-quality spectra for the faint galaxies in A\,2218
at $z=0.18$, we designed a new high dispersion grism optimised for
red wavelengths. The existing LDSS2 high-dispersion blue grism has
relatively low throughput in the wavelength range of interest here
($\lambda\lambda=6000$--7000\AA).  More critically, the undeviated
central wavelength of this grism is 4200\AA, this severely limits the
field of view that can be obtained because of the large deviation of
the redder wavelengths which cause them to fall at the extreme edge of
the detector.

The new high-dispersion red grism, \textsf{R640},  consists of a
600\,l/mm, $34^{\circ}$-blaze angle transmission grating coupled to a
$32.7^{\circ}$ Schott SF10 high refractive index prism. This
combination achieves an undeviated central wavelength of
6500\AA\ allowing the whole of the field-of-view to be used to select
galaxies. The intrinsic dispersion of the SF10 glass also helps boost
the dispersion of the grism by $\approx$8\% providing a measured
dispersion of 2.1\AA\ per pixel with the SITe CCD (24$\mu$m pixels,
spatial resolution $0.59''$\,pixel$^{-1}$).  The peak efficiency of the
grism, $\approx55\%$, represents a gain of almost a factor of two in
the $\lambda\lambda=6000$--7000\AA\ region over that available from
the high-dispersion blue grism.

With a slit width of $1.5''$ the \textsf{R640} grism provides a
limiting spectral resolution element of 4.6\AA\ FWHM. This allows us to
measure velocity dispersions for galaxies as low as $\approx120$\,\kms,
at which point, the intrinsic and instrumental broadening of the
spectra are equal.  Observing with the \textsf{S1\,3750} blocking
filter the total restframe wavelength range covered by the SITe CCD was
$\lambda\lambda\approx4900-6400$\AA\ encompassing the important
absorption lines H$\gamma$, H$\beta$, \mgb\ and Fe5270 at the cluster
redshift.

The three masks were observed with total exposure times of about 5
hours each (Table\,\ref{mask}). Of the 61 galaxy spectra, 12 galaxies
were included on two different masks to allow us to check for
systematic variations between the results from the various masks. There
is only one foreground galaxy (\#2302, only 10,000\,\kms\ away from
A\,2218) which demonstrates the efficiency of our sample selection.  In
total therefore we obtained spectra of 48 different cluster galaxies,
of which 19 lie within the {\it HST} image.  They were morphologically
classified visually by Prof.\ W.\ Couch, who classifies them as 8
E, 1 E/S0, 5 S0, 3 SB0/a, 1 Sa, 1 Sab  
(see Table\,\ref{hstpar}), hence galaxies with disks make up 50\% of our
{\it HST} sample.  In addition five of the galaxies outside the {\it
HST} field have clear evidence for a disk component (see
Table\,\ref{linpar1}), but the modest seeing ($0.95''$ in $I$) of our
ground-based images prevents us from classifying them in more detail.

\begin{table}
\centering
\begin{minipage}{60mm}
\caption{Observation of the masks.}
\label{mask}
\begin{tabular}{lcc}
Mask & number of galaxies & $t_{\mathrm{exp}}$ [hr] \\
\hline
2 & 24 &  5.50 \\
5 & 19 &  4.75 \\
7 & 18 &  4.50 \\
\end{tabular}
%
\end{minipage}
\end{table}

\subsubsection{Reduction and analysis of the spectra}
\label{specredu}

The spectral reduction was undertaken using {\sc midas} with own {\sc
fortran} routines and followed the standard procedure, including
correction for  S--distortions of the spectra closest to the edges of
the field. The individual 2-d images of the slitlets were
extracted from the whole frame (after bias subtraction) and reduced
individually. Dome flat--fields were used to correct for pixel-to-pixel
variation, cosmic rays were removed by a $\kappa$--$\sigma$ clipping
algorithm with a $5\times 5$ pixel filter and bad columns were cleaned
by interpolating adjacent columns (there were 0 to 5 bad columns per
slitlet).  The rectification was achieved by tracing the spectral
profiles and then shifting the pixels in the spatial direction. This
transformation was applied in the same manner to the science,
calibration and sky flat images. After sky flat--fielding, to correct
for illumination effects, the spectra were wavelength calibrated and
the sky was subtracted by modeling each CCD column seperately.
One-dimensional spectra were extracted using the Horne--algorithm
\cite{Horne86}, which optimally weights the extracted profile to
maximise the signal-to-noise.  Finally, the one-dimensional spectra
were rebinned to logarithmic wavelength steps in preparation for the
{\it Fourier Correlation Quotient} (FCQ) determination of the velocity
dispersion and the measurement of absorption line strengths.

Spectra of standard stars were reduced in a similar manner. A
spectrophotometric flux standard (BD$+$28\,4211) was observed through
an acquisition star hole in one mask. Template G and K giants stars
(HD\,102494, HD\,107328, HD\,126778, HD\,132737 and HD\,184275) were
observed through a longslit using the same grism as for the
galaxies. To minimize the effect of the possible variation in slit
width the star spectra were summed over a small number of rows.

The velocity dispersions (as well as the radial velocities) were
determined using the latest version of the FCQ program kindly provided
by Prof.\ R.\ Bender (see \pcite{Bende90a}). The wavelength range
analyzed ($\lambda\lambda=5943$--6279\AA) was centered on the \mgb\
feature and lies between two very strong telluric emission lines. The
resulting dispersions cannot be interpreted simply, since the slitlets
had small variations in width and were therefore not identical in width
to the longslit. We applied a procedure to corrrect for this. FCQ was
run on all galaxies to give a $\sigma_{{\rm fcq}}$ for each of the five
template stars. Separately we computed the width of the
auto-correlation function of each star, $\sigma_{\rm star}$, and added
this in quadrature to estimate the total line width of the galaxy
absorption lines using each template : $\sigma_{{\rm tot}} =
\sqrt{\sigma_{{\rm fcq}}^2+\sigma_{{\rm star}}^2}$. We determined the
final value of the stellar velocity dispersion by taking the median
value of these five measurements, $\overline{\sigma_{\rm tot}}$, and
subtracting from it (in quadrature) the instrumental dispersion,
$\sigma_{{\rm inst}}$ i.e.\  $\sigma_{\rm gal} =
\sqrt{\overline{\sigma_{\rm tot}}^2 - {\sigma_{\rm inst}}^2}$. The
instrumental dispersion is determined using eight unblended emission
lines in the arc spectrum, from the appropriate slitlet, to
evaluate the width of an arc line at the position of \mgb\ in the
redshifted galaxy spectrum. All the values of $\sigma_{{\rm gal}}$ are
given in Table\,\ref{linpar1} together with the heliocentric radial
velocities $v_{{\rm rad}}$.

For comparison, velocity dispersions of all galaxies of one mask were also
measured using the {\sc iraf} package {\sc fxcor} yielding an
average difference of only 28\,\kms.  We also compare our velocity
dispersion results with those of early-type
galaxies in common with JFHD (see Table\,\ref{cfsig}).  The
median difference in $\sigma_{{\rm gal}}$ is 22\,\kms\ or 10\%.

\begin{table*}
\centering
\begin{minipage}{95mm}
\caption{Comparison between our measurements and JFHD.}
\label{cfsig}
\begin{tabular}{lrrrrrr}
Galaxy & $v_{{\rm rad}}$ & $v_{{\rm JFHD}}$ & $\Delta v$ & 
$\sigma_{{\rm gal}}$ & $\sigma_{{\rm JFHD}}$ &  $\Delta\sigma$  \\
\hline
1293 & 47895.8 & 47882.5 & 13.3 & 239.9 & 201.4 & 38.5  \\
1343 & 50026.9 & 49986.2 & 40.7 & 241.5 & 248.3 & $-$6.8  \\
1437 & 48495.0 & 48478.5 & 16.5 & 248.5 & 204.6 & 43.8  \\
1662 & 44948.2 & 44928.6 & 19.5 & 364.9 & 298.5 & 66.4  \\
1711 & 47619.0 & 47608.8 & 10.2 & 203.7 & 215.3 & $-$11.6  \\
1914 & 47555.5 & 47534.1 & 21.4 & 147.4 & 162.2 & $-$14.7  \\
2076 & 49399.4 & 49394.3 &  5.1 & 219.0 & 196.8 & 22.2  \\
2604 & 49390.2 & 49418.9 & $-$28.8 & 187.8 & 130.9 & 56.8  \\
\end{tabular}

\medskip
Galaxy numbers correspond to those in Table\,\ref{linpar1},
$v_{{\rm rad}}$ heliocentric radial velocity
measured by us,
$v_{{\rm JFHD}}$ heliocentric radial velocity
measured by JFHD,
$\Delta v$ the difference between both measurements.
$\sigma_{{\rm gal}}$: velocity dispersion measured by us and corrected to
the same aperture used by JFHD,
$\sigma_{{\rm JFHD}}$: velocity dispersion measured by JFHD,
$\Delta\sigma$ the difference between both measurements.
All velocities are in \kms.
\end{minipage}
\end{table*}

The absorption indices were measured on the Lick system
\cite{FFBG85}. Our spectra were first degraded to
the resolution of the Lick system. The indices were then corrected for the
velocity broadening $\sigma_{{\rm gal}}$ (a few example spectra are shown in
Fig.\,\ref{spectrum}).  Because the equivalent
widths are determined with respect to a local continuum, flux
calibration has little effect on the (atomic) absorption line strengths
(the average difference in Mg$_b$ for example is $-0.014$\AA, which is
about 10\% of the typical error). Nevertheless, we present in
Table\,\ref{linpar1} the line strengths of H$\beta$ and Mg$_b$
measured from the flux-calibrated spectra (values of other indices may be 
requested from the first author). 
We calibrated the data using standard stars in common
with the EFAR studies \cite{CBDMSW99}. These showed a scatter around zero with
$\sigma=\pm0.1$\AA\ for \mgb\ and hence no recalibration was applied to
the index strengths to place them on the Lick system.

The absorption of
the iron lines Fe5270 and Fe5335 could not be  derived reliably for the
majority of the galaxies because the red continuum band of Fe5270 is
contaminated the [OI] telluric emission line at
$\lambda=6296$\AA.  The Fe5335 line is also redshifted into 
this sky line, so that this absorption feature cannot be measured.  The
same problem also arises for the red continuum window of  Mg$_2$.

\begin{figure*}
\centerline{\psfig{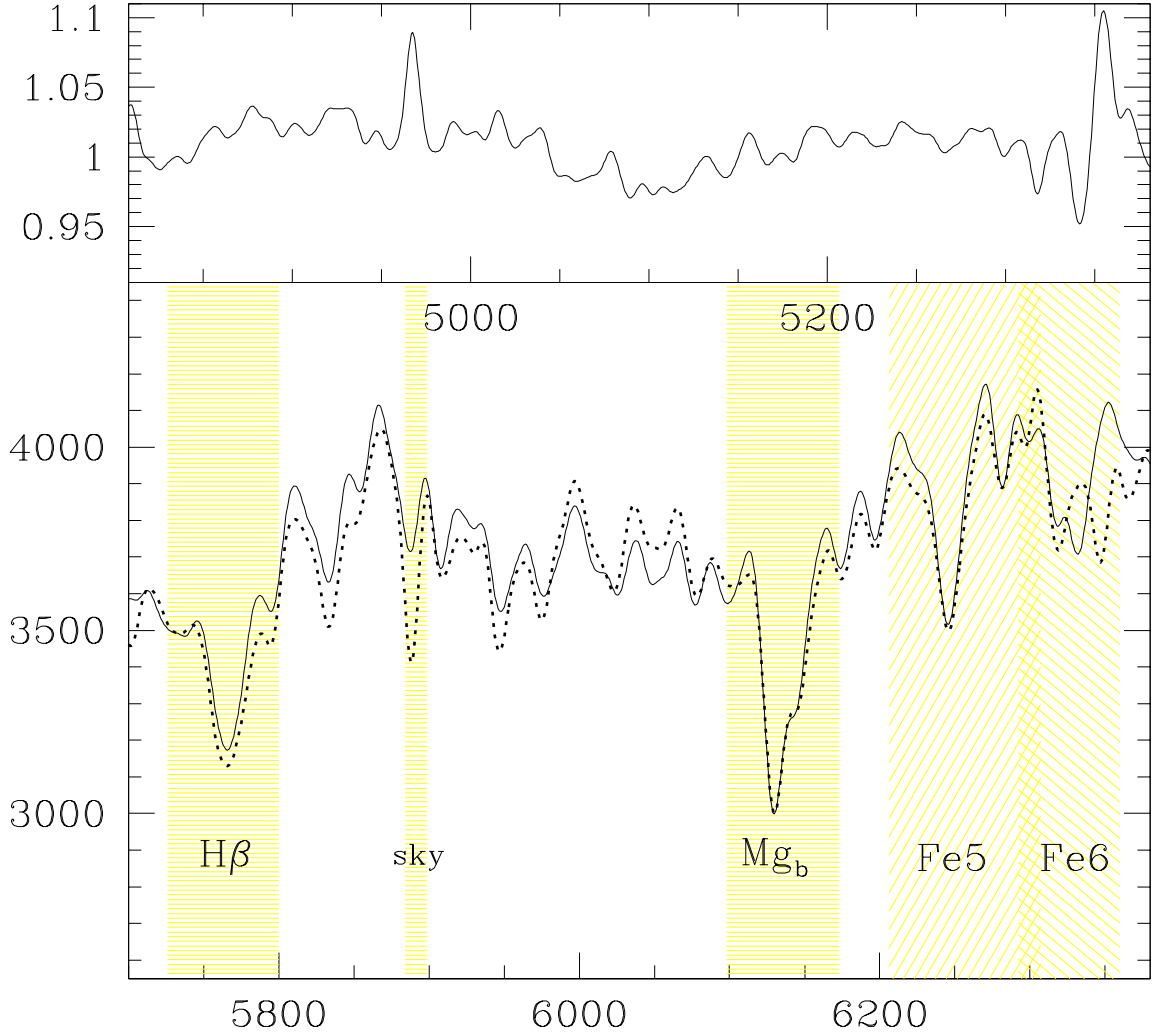}}
\caption{Lower panel:
  Flux-calibrated spectra of galaxy \#786 observed through two different
  masks. The spectra have been degraded to the Lick resolution and normalized
  at \mgb. Index definitions are overplotted as hatched areas (Fe5=Fe5270, 
  Fe6=Fe5335). The continuum as well as the \hb\ and \mgb\ indices are
  all well matched. 
  Only the Fe5335 index is severely affected by two sky 
  emission lines. The lower axis represents
  observed wavelengths, the upper shows restframe wavelengths (in \AA).
  (See also Fig.\,\ref{spectrum}). The abscissa is in counts (1\,ADU $=1.3e^-$).
  Upper panel: ratio of the 2 spectra.}
\label{compspec}
\end{figure*}

~From  the repeat observation of 12 galaxies on different masks we are
able to confirm the internal reliability of our spectral analysis. As
an example we show two independent spectra of galaxy \#786 in Fig.\,\ref{compspec},
indicating that the continuum as well as the H$\beta$ and Mg$_b$ 
indices are well matched.  If we take all 12 comparison spectra
regardless of their signal-to-noise we find median offsets for the
H$\beta$ and Mg$_b$ line indices and  velocity dispersion,
$\sigma$, at  the $<10$\% level.

\section{The Local Comparison Sample}

\subsection{Photometry and Structural Parameters}

The Coma cluster provides a good local reference as it represents the
best studied rich local cluster, albeit less rich than A\,2218.  We
take data on early-type galaxies in the cluster from \scite{Joerg99}
and \scite{JFK95b}.  Their galaxy photometry was taken through the Gunn
$r$ filter, then corrected for extinction and cosmic expansion.
The combined sample
contains 115 early-type galaxies (35 E, 55 S0, 25 intermediate types)
and is 93\,\% complete at absolute magnitudes $M_r<-20.35$.

\subsection{Velocity Dispersions}

The spectroscopic data in \scite{Joerg99} and \scite{JFK95} were
aperture corrected by the authors to match a circular aperture with
radius 1.7$''$. Therefore, we corrected our velocity dispersion
($\sigma$) determinations by $\Delta(\log \sigma) = +0.025$ to match
the Coma aperture.  The aperture correction was computed from the logarithmic 
gradient given by \scite{JFK95}:
\begin{equation}
\label{sigapcor}
\Delta(\log \sigma) = 0.04\cdot
\log\left(\frac{d({\rm A\,2218})}{d({\rm Coma})} \cdot 
\frac{a({\rm A\,2218})}{a({\rm Coma})}\right)
\end{equation}
where $d$ is the angular distance and $a$ is the aperture radius. 
For our observations $a$ was taken as the harmonic mean of the slit width
of 1.7$''$ and the weighted number of rows (an average of 4.7 pixels, 
or 2.8$''$) over which the spectra were integrated by the Horne algorithm.

\subsection{Line Indices}

The best available sample of line indices for local early-type galaxies comes 
from the SMAC
collaboration \cite{KLSHD00}. We use this dataset therefore as the
primary comparison source for our line index analyses.  While the SMAC sample
includes galaxies from lower density regions (such as the Virgo cluster),
about half of the dataset comes from the Coma cluster.  
We chose not to use line indices from \scite{Joerg99} because the
H$\beta$ absorption strengths are not compatible with the higher S/N
observations of \scite{KLSHD00} (see Fig.\,1 of that paper).

The Mg equivalent widths of the A\,2218 galaxies were aperture
corrected according to the same prescription given by \scite{JFK95} as for the
velocity dispersions. For the conversion between Mg$_2$ and \mgb\ we
follow \scite{ZB97} and adopt a factor of 15. This leads to a coefficient of 0.6
(instead of 0.04) in Eq.\,\ref{sigapcor} resulting in a mean aperture 
correction of $\Delta({\rm Mg}_b) = +0.37$\AA\ for the A\,2218 galaxies.
We did not correct the \hb\ line
strengths for aperture effects since no significant radial dependence
of this index is found in local galaxies \cite{MSBW00}.

\section{Results}

\subsection{Comparison with local data}

We begin by explaining the symbols used in the figures on which the
following discussion is based.  In all subsequent figures, large
symbols are galaxies in A\,2218, while small boxes represent the local
reference sample. Galaxies classified morphologically as S0 or early-type spiral
galaxies (from the A\,2218 {\it HST} field) as well as five galaxies that clearly exhibit disks in the ground-based
images are shown by filled symbols. In Section\,\ref{raddep}, we will divide the full A\,2218
sample radially into two subsamples with equal numbers of galaxies.
The core region (all galaxies within 130\arcsec\ from the cluster
centre) overlaps with the {\it HST} field (providing morphological
classifications) and  we show these galaxies as circles.  Galaxies in
the outer region (triangles) may contain small disks which can not be
detected on the ground-based images.  We fit both the distant and the
local sample only within the region shown by horizontal and/or vertical
dotted lines in the plots, which represent the selection boundaries for
the A\,2218 data.

For linear fits to the relations we use the bisector method, which is a 
combination of two least-square fits with the dependent and independent 
variable interchanged. Errors on the bisector fits were determined by
bootstrap resampling the data 100 times. The shaded area in each figure
illustrates the possible
slopes within the $\pm1\sigma$ bounds on the mean slope of the A\,2218 sample,
whereas the two solid lines indicate the same range for the local comparison
data. All fit results are given in Table\,\ref{fitpar} in the Appendix.

\subsubsection{Faber--Jackson relation}
\label{fjr}

\begin{figure*}
\centerline{\psfig{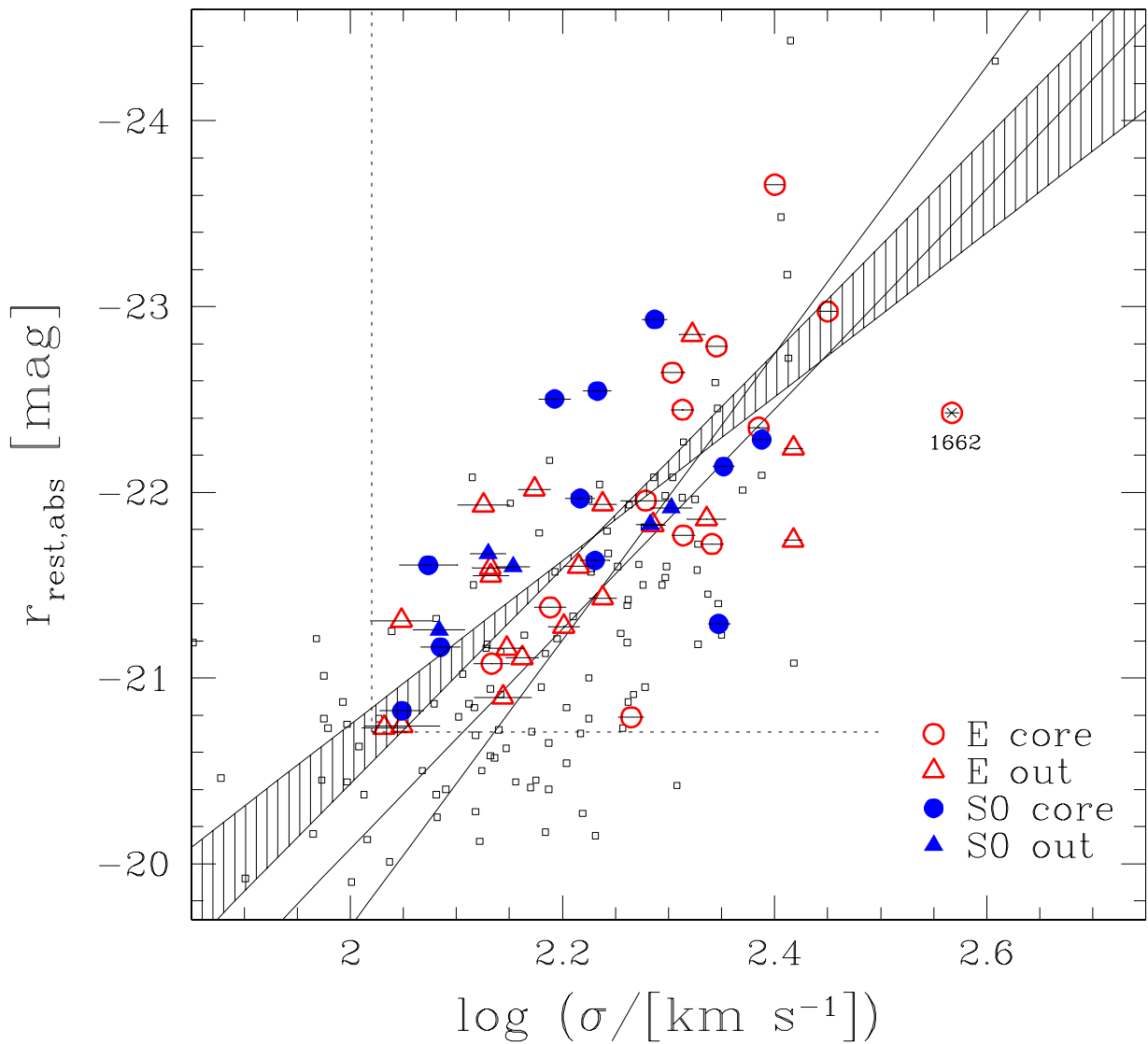}}
\caption{The Faber--Jackson relation (velocity dispersion {\it versus} total 
restframe Gunn $r$ magnitude) for the galaxies in A\,2218 (large symbols), 
compared to Coma early-type galaxies  
(small open boxes; J{\o}rgensen et al. 1995, 1999).
Hashed area: fits to the distant FJR (within $\pm 1\sigma$),
open area: fits to the local FJR within selection boundaries of the
distant sample (dotted lines).}
\label{sigbabs}
\end{figure*}

The Faber--Jackson relation in A\,2218 is compared to the Coma sample of
J{\o}rgensen et al.\ (1995b) in Fig.\,\ref{sigbabs}. 
Due to the lookback time to A\,2218, we should expect galaxies in this
cluster to be brighter than their  counterparts in Coma at a given velocity
dispersion. If we assume that cluster galaxies are formed at $z_f=2$, passive
evolution would increase their brightness by $0.19$\,mag \cite{BC93}.
The observed change in absolute magnitude is
$0.30\pm0.09$\,mag, which is compatible with the theoretical prediction.

With 48 galaxies in A\,2218  covering velocity dispersions down to
$105$\,\kms, we are also able to investigate the evolution
in the slope of the FJR. The slope of the Coma data is steeper than that
in A\,2218 (a difference of $1.7\pm1.1$) but the offset has low statistical
significance, and depends on the boundaries used to define
the Coma galaxy sample.  Thus there is only weak evidence from this plot for
differential evolution of massive and less-massive early-type
galaxies. It is worth noting, however, that the A\,2218 data may also have
larger scatter than that of the Coma sample.  This may indicate a
wider range of recent star formation histories of galaxies in the
A\,2218 cluster, an issue to which we will return in the following
sections.

Looking at the individual galaxies in A\,2218, one in particular stands
out: \#1662 (marked with a cross in Fig.\,\ref{sigbabs}).  This galaxy
has a higher velocity dispersion than the cD galaxy (as measured by
JFHD), it is located close to the cD and shows moderately high
ellipticity, a positive a4 coefficient and some asymmetry in the {\it
HST} image (see Fig.\,\ref{mosaichst}). We confirmed that there was no
other galaxies on the slit and  splitting the spectroscopic exposures
into two independent halves we find roughly equal and high $\sigma$'s.
The origin of this galaxy is puzzling: it could perhaps be the stripped
core of a massive galaxy, although it is hard to understand why the
parent galaxy would have been susceptible to stripping if it were as
large as suggested by the dispersion.

\subsubsection{Line index analysis}

\begin{figure*}
\centerline{\psfig{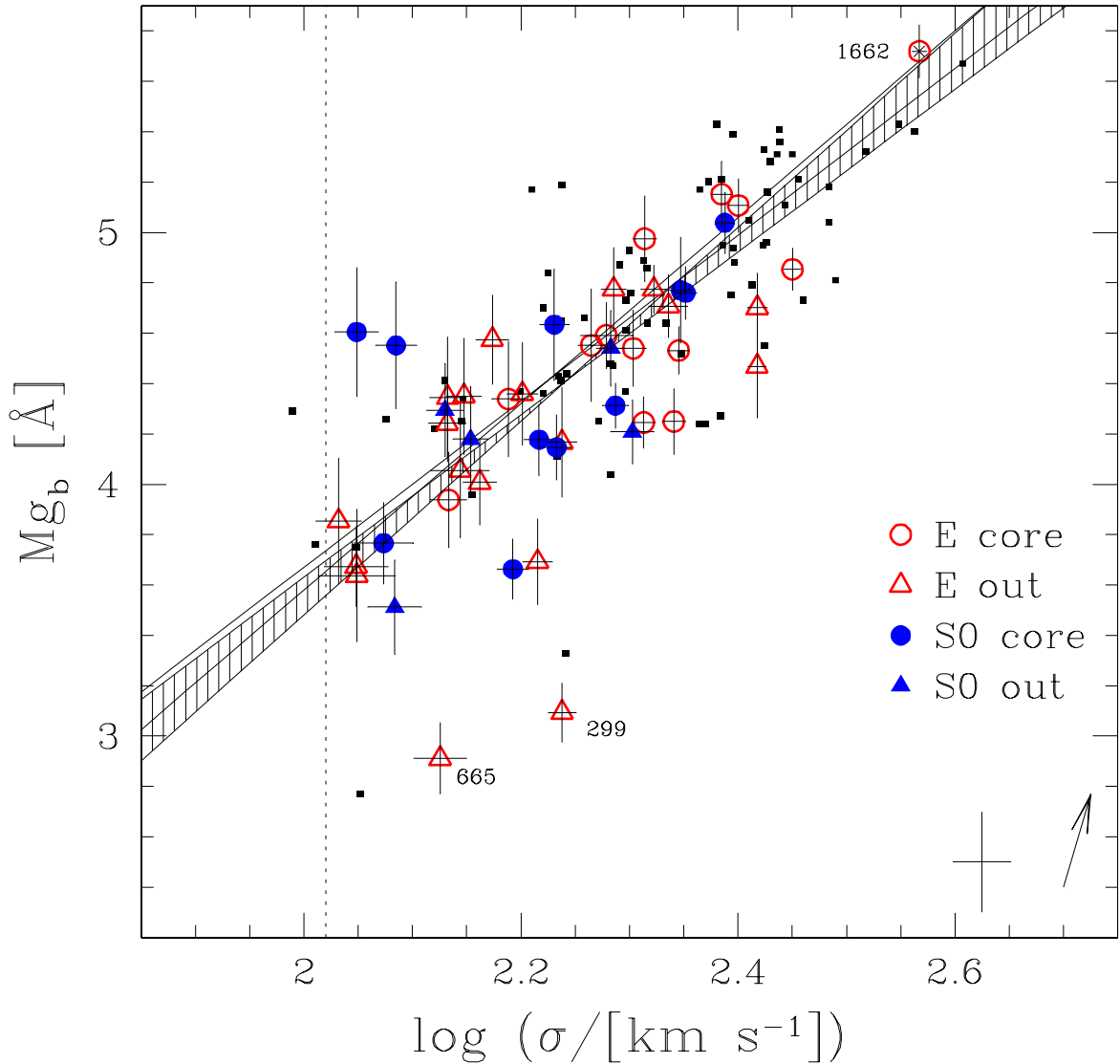}}
\caption{Mg$_b$--$\sigma$ relation of A\,2218 compared to the SMAC sample
(small filled boxes).
Symbols and lines as in Fig.\,\ref{sigbabs}.
The arrow in the lower right corner shows the aperture
correction applied to the measured $\sigma$ and Mg$_b$ of the distant 
galaxies, the cross
represents the median errors for the SMAC data (Kuntschner et al.\ 2000).}
\label{sigmgb}
\end{figure*}

Overall, the distribution of A\,2218 galaxies in the \mgb--$\sigma$ plane is
quite similar to that of local early--type galaxies in the SMAC sample
(see Fig.\,\ref{sigmgb}). However, two galaxies in A\,2218 (\#299 and \#665) 
stand out as peculiar, having
low \mgb\ line strengths for their $\sigma$'s. \#299 has a strong
\hb\ absorption so that it may be a post-starburst (k+a or \ea) galaxy,
while \#665 has an average \hb\ value. It is not possible to detect any
stellar disk in either \#299 or \#665 in the ground--based images, but
both galaxies are very blue in ($U-V$) compared to the
luminosity-corrected mean colour--magnitude relation of the total
sample (this is also true for \#704 and \#1605 which have high
\hb\ absorption) perhaps indicating some recent star formation
activity.  The galaxy with the highest $\sigma$ (\#1662, marked with a
cross in Fig.\,\ref{sigmgb}) has a \mgb\ line strength which is compatible
with the \mgb--$\sigma$ relation of the other A\,2218 galaxies.

Whether or not we excluded the galaxies \#299 and \#665 from the bootstrap
bisector fits, the slopes for the A\,2218 and local samples are
consistent ($\Delta\hbox{slope}=0.18\pm0.58$ with these galaxies
included; $\Delta\hbox{slope}=-0.26\pm0.48$ with these galaxies excluded).
The offset between the two samples is more dependent on whether these 
galaxies are included, since they tend to reduce the average \mgb\
absorption ($\Delta\hbox{zp} = -0.061\pm0.074$\AA\ with the galaxies
included; $\Delta\hbox{zp} = -0.004\pm0.072$\AA\ with the galaxies excluded).

We can compare this shift with that expected due to the passive
evolution of the galaxy population by adopting a formation redshift.
For $z_f=2$, we expect a change $\Delta\hbox{zp} = -0.19$\AA\ (assuming
a typical
[Fe/H] of 0.25 for the sample, see \pcite{ZB97}). 
Since we see a significantly smaller offset,
the average galaxy must have been formed at 
a considerably earlier time. However, if we exclude the galaxies \#299 
and \#665 we may well be removing exactly those galaxies which are 
showing evolutionary differences. Including these galaxies and allowing
for a systematic offset of 0.1 in the relative calibration of the 
spectral index
would reduce the difference to within the statistical uncertainty of
our fiducial model.

\begin{figure*}
\centerline{\psfig{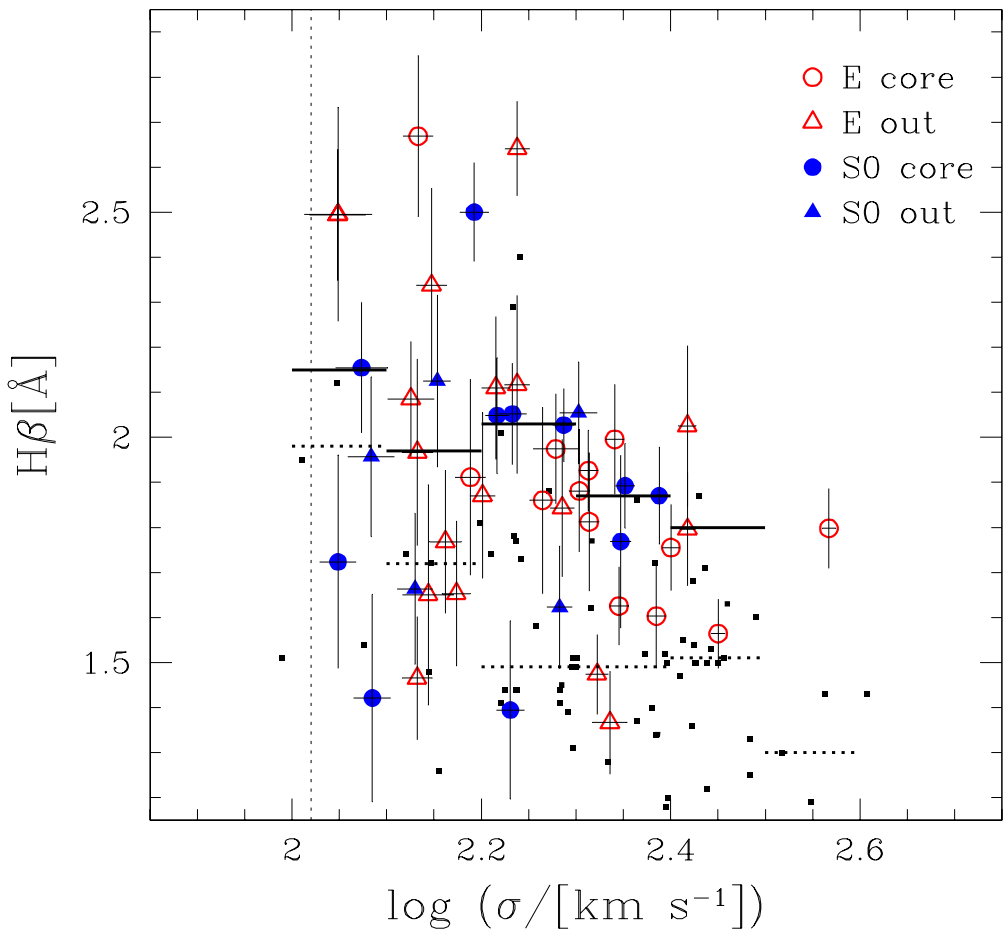}}
\caption{The distribution of the H$\beta$ line indices of the A\,2218
early--type galaxies compared to SMAC. Symbols as in
Fig.\,\ref{sigbabs}.
Short horizontal solid lines: median value of \hb\ for specific $\sigma$ bins
for A\,2218; horizontal dotted lines: same for SMAC.}
\label{sighb}
\end{figure*}

\begin{figure*}
\centerline{\psfig{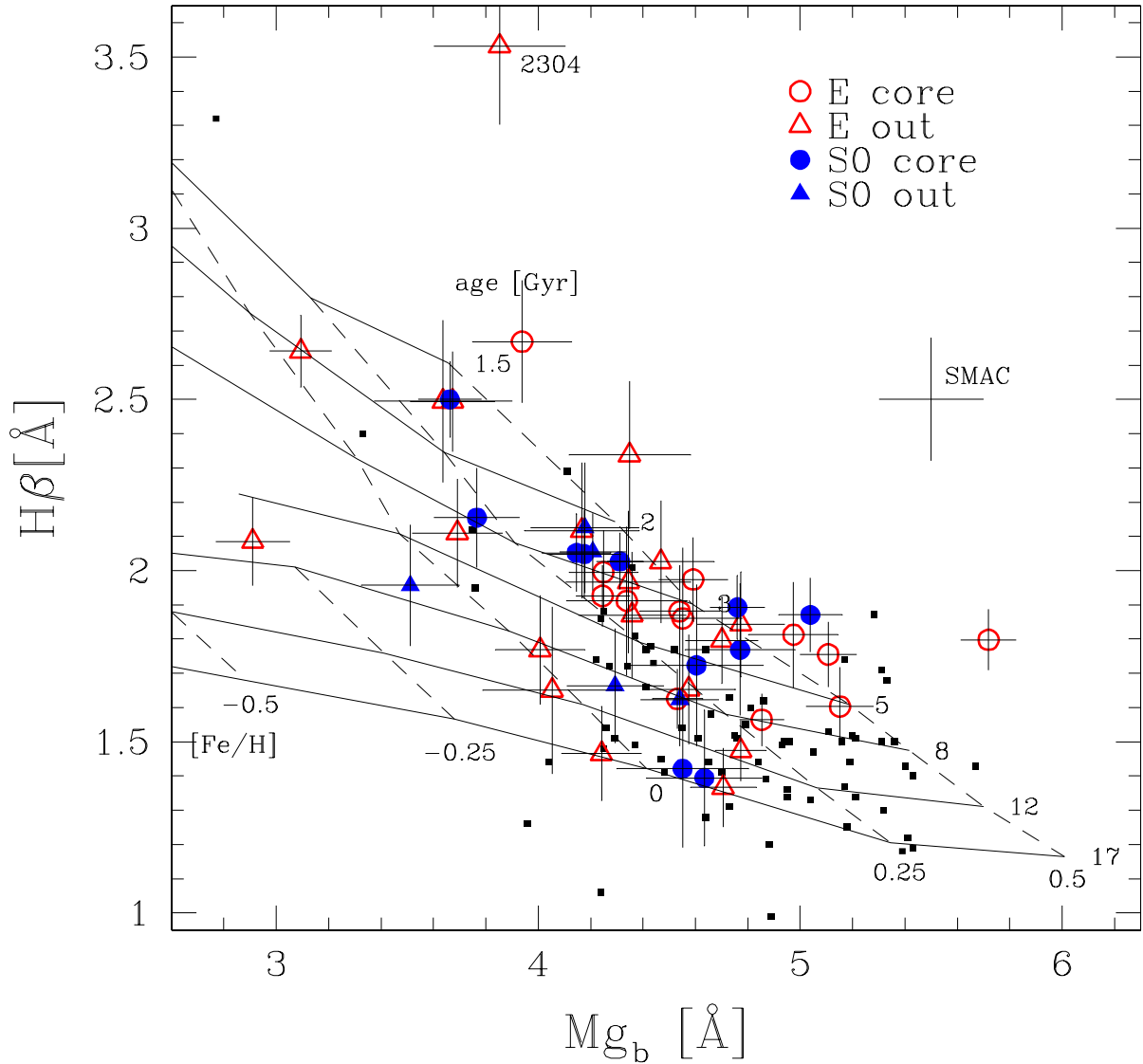}}
\caption{Age -- metallicity diagnostic diagram using H$\beta$ as a
primary age indicator and \mgb\ as a metallicity indicator. Big symbols: 
A\,2218, small filled boxes: SMAC sample (Kuntschner et al.\ 2000, typical 
errorbar shown). Overplotted is a grid of SSP models from
Worthey (1994) shifted as described in the text. ``Horizontal'' lines
follow constant age, ``vertical'' lines constant metallicity.
}
\label{mghb}
\end{figure*}

Since the \hb\ line index is more age sensitive, but less dependent on
metallicity than \mgb, we explore the age/metallicity spread of the
A\,2218 galaxies in more detail in Figs.\,\ref{sighb} and \ref{mghb}.
The distribution within the \hb--$\sigma$ plane is broad. In particular,
the large scatter of the fainter galaxies in \hb\ indicates a wide range of
star formation histories. The more massive galaxies in A\,2218 have on
average higher \hb\ absorption than galaxies in the SMAC sample with
the same velocity dispersion. This is illustrated in Figs.\,\ref{sighb} by
calculating the median \hb\ aborption in five bins 
(medians for each sample are indicated by solid
and dotted horizontal bars respectively in Fig.\,\ref{sighb}). Although
there is a trend for $H\beta$ to increase with decreasing velocity
dispersion, this effect is complicated by the systematic variation
in metal abundance with velocity dispersion. We return to this issue
below, and firstly concentrate on the offset of the brighter galaxies.
Comparing the four brightest bins, the offset is $\approx0.35\pm0.1$\AA.
We are confident that this systematic offset cannot be produced by 
calibration errors
since our systematic errors are less than 0.1\AA.

In order to interpret this difference, we compare the absolute
values of the \hb\ index with the Worthey model (1994), and also
in a relative sense only. The absolute values of
the index suggests that galaxies in the SMAC sample have 
luminosity weighted ages for their stellar populations of 8\,Gyr,
hile A\,2218 galaxies are much younger, 
with luminosity weighted ages of only 4\,Gyr (compare with Fig.\,\ref{mghb}). 
However,  the absolute
calibration of the model is uncertain. A more reliable approach is 
to compare the difference in absorption line strength with that expected
from the difference in lookback time.
The expected difference in \hb\ line strength for a single population
formed at $z_f=2$, observed at $z=0.17$ and the present day is
0.14\,\AA.   This is less than the observed shift, although   
the two values could be reconciled if we include both
the maximum systematic uncertainty of 0.1\,\AA\ in the index calibration 
and the statistical uncertainty.  Nevertheless, it is interesting to note 
that the pressure from
the \hb\ index is to introduce more recent star formation, while the 
pressure from the \mgb\ observations is to make the galaxies older. It is 
unlikely that this conflict arises from our
use of a single age stellar population to interpret the
differences as the intermediate
age population will have similar effects on both indices. 


Exploring the distribution of A\,2218 galaxies between age and
metallicity using line diagnostic diagrams has the advantage over
the \mgb--$\sigma$ and \hb--$\sigma$ diagrams, in that it does not rely
on assuming a universal metal abundance -- velocity dispersion relation
before we can extract any information on the ages of the stellar
populations.  We use the \mgb--\hb\ diagram, Fig.\,\ref{mghb}, rather than 
the combined [MgFe] index, because we were
unable to derive reliable Fe indices for all our galaxies. The
\mgb--\hb\ correlation has the drawback that the models based on solar
abundance ratios do not match the Mg/Fe ratio observed in early-type
galaxies \cite{TGB99}.  However, we are primarily interested in the
relative ages of galaxies, so  we can apply an empirical relationship
between \mgb\ and [MgFe] from the SMAC data to the model grid of
\scite{Worth94}: \mgb$=$[MgFe]$*1.58-1.20$, to transform it to the
relevant observables. While ages derived from this grid are clearly
uncertain in an absolute sense, this approach allows us to quantify the
relative offsets between the data.

The bulk of the
A\,2218 galaxy population show systematically younger ages at a fixed
metallicity compared to the SMAC galaxies (Fig.\,\ref{mghb}). 
This contributes to a
deficit of A\,2218 galaxies in the lower right-hand corner of the plot
compared to the SMAC sample. There is also a systematic shift between the
two samples that is driven by the \hb\ index. The magnitude of the
offset is larger
than that expected for the look-back time to A\,2218 if the galaxies
form all their stars at redshifts above 2, as we have discussed
previously. Since the velocity dispersions of the galaxies are not 
directly visible in
this diagram, the comparison suggests that the evolution of \mgb\ and \hb\ 
are consistent; it only becomes evident from comparing the
distribution of galaxies in Figures \ref{sigmgb} and~\ref{sighb} that
the SMAC sample contains a higher proportion of high velocity
dispersion galaxies. Thus, while
galaxies in A\,2218 occupy a large region of the available parameter
space in Fig.\,\ref{mghb} and appear to be more diverse than the SMAC
population, the result is not conclusive because low $\sigma$ galaxies,
which in general show a wider range in \hb,
 are under-represented in the SMAC sample. 

A few galaxies in A\,2218 are exceptions to the general distribution, 
having \hb\ as low as galaxies in the SMAC dataset. We have searched for 
emission in these
galaxies, which might fill-in the \hb\ line, but find none.
In contrast, galaxy \#2304 has very strong Balmer absorption 
in \hb\ and also in the \hg\ and \hga\ indices. This galaxy has
clearly undergone a starburst in the recent past, and is an example of
the post-starburst galaxies frequently identified in
lower-resolution studies of more distant clusters (e.g.\ \pcite{PSDCB99}).

\subsubsection{Fundamental Plane}
\label{secfp}

\begin{figure*}
\centerline{\psfig{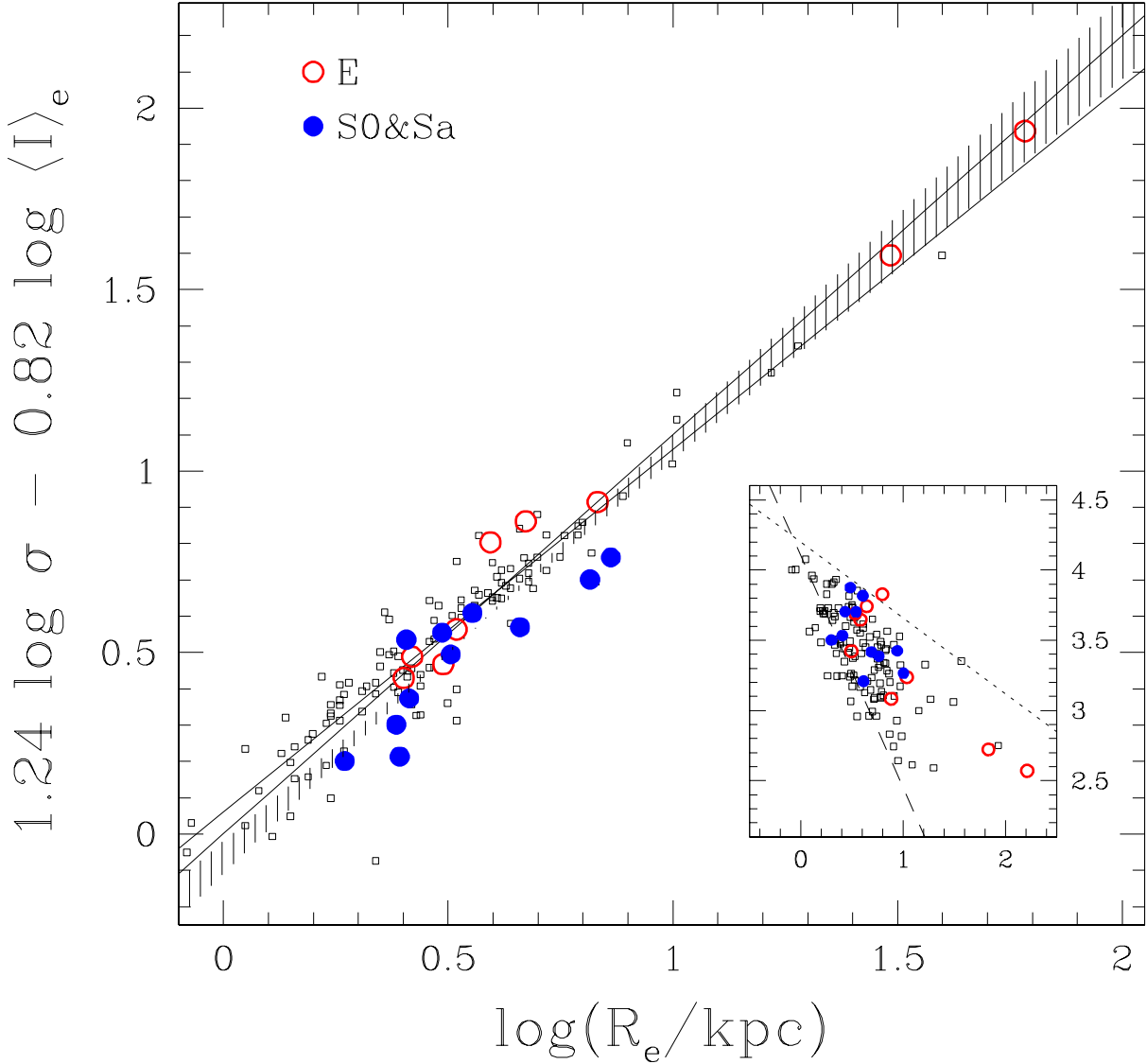}}
\caption{The Fundamental Plane of A\,2218 early-type galaxies (big symbols)
compared to Coma (small open boxes, J{\o}rgensen et al.\ 1995a; 
J{\o}rgensen 1999). Lines as in Fig.\,\ref{sigbabs}.
The {\it HST} F702W photometry was transformed to restframe
Gunn $r_{rest}$ magnitudes as described in Section\,\ref{rfsbp}.  
The full diagram shows the edge-on view, while the inset
 shows the FP face-on (using the same symbols).  The abscissa is
$(2.21\log(R_e)-0.82\log\langle I\rangle_e+1.24\log\sigma)/2.66$,
ordinate $(1.24\log\langle I\rangle_e+0.82\log\sigma)/1.49$.  The
dotted line indicates the ``exclusion zone'' for nearby galaxies, the
dashed line the magnitude completeness limit for the Coma sample.}
\label{fp}
\end{figure*}

Another probe of the star formation histories of galaxies is the stellar
mass-to-light ratio.
We can investigate the Fundamental Plane (FP) of A\,2218, restricting
the analysis to those galaxies within the {\it HST} field where
accurate measurements of the structural parameters are possible. In
Fig.\,\ref{fp}, we present the FP and compare the distant
galaxies to the Coma sample of J{\o}rgensen et al.\ (1995a, 1996).
The  scatter around the
FP of A\,2218 is with 0.108 in $\log(R_e)$.  This is very similar to the
scatter seen in
Coma, 0.096, as is the distribution of the galaxies across the
surface of the plane (inset panel in Fig.\,\ref{fp}). 
The average brightening of the A\,2218 early-type
galaxies can be investigated by assuming that there is no evolution 
in the structure of the galaxies (i.e.\ $R_e$ and $\sigma$ remain fixed).
In this case the evolution of the zero-point of the relation
measures the increase in brightness. The zero-point offset is
$0.067\pm0.023$\,mag, consistent with a $z_f=2$ passive evolution model 
which predicts a shift of 0.10\,mag. 

\begin{figure*}
\centerline{\psfig{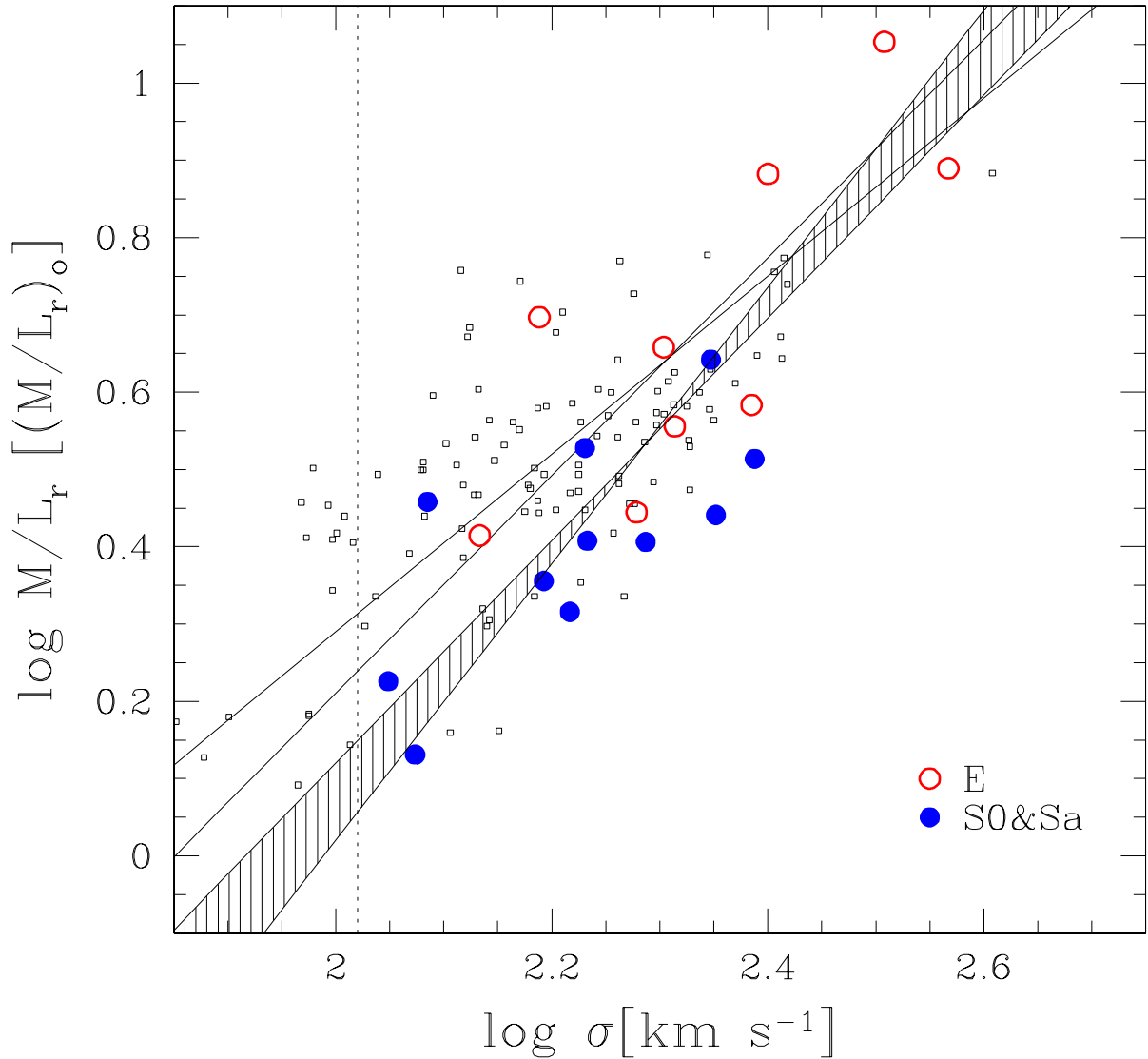}}
\caption{The mass-to-light ratio versus velocity dispersion. 
Symbols as in Fig.\,\ref{fp}.
Hashed area: fits to the A\,2218 galaxies (within $\pm 1\sigma$),
open area: fits to the Coma sample within selection boundaries of the
distant sample (dotted line).}
\label{sigml}
\end{figure*}

To cast this discussion in terms of mass-to-light ratios
(Fig.\,\ref{sigml}) we calculated the masses of the
galaxies following JFHD: $M=5R_e\sigma^2/G$,  based on 
\scite{BBF92}. \footnote{
For the absolute magnitude in Gunn $r$ of the Sun, we take
$M_{r,o}=4.87$\,mag which was derived from $M_{V,o}=5.72$ and
$(V-R)_o=0.52$ \cite{LB} and the transformation $r=R+0.41+0.21(V-R)$
\cite{Kent85}.} We limited the Coma sample to galaxies with
$\log\sigma\geq 2.02$ (dotted line in Fig.\,\ref{sigml}) in order to
match the area of parameter space covered by the A\,2218 galaxies.
The bootstrap bisector fits to $M/L = a \sigma^m$ show compatible M/L
slopes ($\Delta m = 0.34\pm0.20$) but a systematic offset in
the zero point of the relation, $\Delta a = 0.080\pm0.026$. This 
is consistent with the expected change due to passive evolution:
$\Delta a = 0.12$.

In their $M/L$ analysis, JFHD find a substantial evolution in the 
slope of this relation. 
While their Coma value ($m=0.66\pm0.13$) is much flatter than the slope
for their five intermediate redshift clusters ($m=1.49\pm0.29$ including A\,2218), the
slopes we determine for Coma and A\,2218 are both compatible (and 
similar to that found for  their
intermediate redshift clusters by JFHD).
The difference between our estimate of the Coma slope and that
from JFHD must result from the different fitting 
methods applied,
in particular our exclusion of the lowest velocity dispersion galaxies.

Finally, we note that we have 
searched for a trend in \hb\ line strength with M/L, but did not find one. 
Thus the naive  expectation that a recent star formation event would
equally effect a
galaxy's brightness and its Balmer line strength is not apparent in our
dataset.

\subsubsection{Summary and Discussion}

The overall result from the comparison of bright galaxies in A\,2218
and the local reference cluster, Coma, is that the galaxy population has
evolved little between $z=0.18$ and the present-day. This is consistent
with an early formation epoch for the bulk of their stellar
populations.  The distribution of the galaxies in the \mgb--$\sigma$
plane and on the Faber--Jackson relation is almost identical in the two
clusters.  Weak evolution of bright galaxies in clusters sets strong
limits on their formation epoch, as has been discussed extensively for
the Mg--$\sigma$ relation \cite{ZB97}  and the Fundamental Plane at
intermediate redshifts (e.g.\ \pcite{BSZBBGH97b,JFHD99}), and even out
to $z\sim 1$ (e.g.\ \pcite{DFKI98}). 
The changes we see in the dynamical relations for bright
galaxies reinforces the conclusions of these authors. However, we are
also in a position to address the ages of stellar populations directly
using the age sensitive \hb\ line index.

In the \hb--$\sigma$ diagram (Fig~\ref{sighb}), the distant galaxies cover a wide range
in \hb, indicative of differences in the stellar populations.  The
low-dispersion galaxies have the greatest range of \hb\ line
strengths. Some of these galaxies clearly had extended star formation
activity at a low level, or recent episodes of minor star formation
involving a few per cent of the total mass. Although the expected increase
in luminosity is less visible in the FJ and FP relations, this is in line
with the study of \scite{ZSBBGS99}.  These authors have investigated
the Kormendy relation (the photometric projection of the FP) of five
distant clusters and concluded that current accuracy of photometric
data (even from {\it HST}) is not high enough to exclude low-level
star formation activity in early--type galaxies.

Comparing the \hb--\mgb\ measurements to models, we
find that the bulk of A\,2218 galaxies have younger luminosity-weighted
ages than nearby galaxies, as expected from the look-back time to
$z=0.18$. The lower luminosity galaxies
exhibit a wider range of stellar ages and metallicities. 
This tendency for lower luminosity galaxies to be less homogeneous is
consistent with the expectation of models of the Butcher-Oemler effect
in distant clusters \cite{SEEB98,PSDCB99,KB00,S00}.  
In the scenarios presented by these authors, the blue, star-forming
galaxies responsible for the Butcher-Oemler effect are expected to
undergo significant fading of their stellar populations when star
formation ceases, as well as perhaps suffering more active stripping of
stars by dynamical processes within the clusters
\cite{SEEB98,PSDCB99}.  For these reasons the evolved descendents of
the Butcher-Oemler galaxies will on average be found in the lower
luminosity galaxy population at lower redshifts.  Equivalently, the
lower luminosity galaxies will show a wider range in their previous
star formation histories and our observations tend to support this
suggestion.

\subsection{E vs S0 Comparison}

Based on morphologies determined from {\it HST}/WFPC2 images,
\scite{DOCSE97} have shown that distant rich clusters contain a greater
fraction of spiral galaxies, and a correspondingly smaller fraction of
S0 galaxies, than similar local clusters. Over the redshift range
0--0.5, the S0 fraction decreases from 60\% locally to 10--20\% in the
higher redshift clusters. In contrast, it appears that the elliptical
galaxy fraction remained relatively constant over the last
5\,Gyrs.  The different evolutionary histories of these populations might
be detectable at $z\approx0.2$, even though the properties of E and S0
galaxies are very similar in the nearby Universe
(e.g.\ \pcite{BBF92,BBF93,SBD93}). In an attempt to uncover such
evidence,  \scite{JSC00} have analysed the combined spectra of E and S0
galaxies in three clusters at $z=0.31$. Surprisingly they found no
significant difference between the stellar populations in the two classes
of galaxies. We repeat this test here, exploiting
the fact that our spectra have higher signal-to-noise allowing us to
search for differences between individual galaxies, rather than having
to look at the average galaxy population.

In Figs.~3--8, we have distinguished
between elliptical galaxies and those that have been morphologically
classified as S0 or early--type spirals from the {\it HST} imaging (see
Table\,\ref{hstpar}) or which show a strong disk component clearly
visible in the ground-based images (see Table\,\ref{linpar1}).  We will
now discuss the differences between these two populations (in the following 
refered to as \lq ellipticals\rq\ or Es and \lq lenticulars\rq\ or S0s,
respectively) as exhibited in those figures.

\subsubsection{Faber--Jackson relation}
We begin by comparing the FJR (Fig.\,\ref{sigbabs}).
Since the E and S0 sub-samples have similar
distributions in $\log\sigma$, the two-dimensional Kolmogorov--Smirnov test 
provides a good measure of the consistency of the two
populations. Applying the KS test yields a probability of
$p=0.66$ that the two distributions are similar. The bootstrap bisector 
fits to the E and S0 sample finds a steeper slope for the S0 galaxies, but
at the $\approx1\sigma$ level (see Table\,\ref{fitpar}). 
There is no significant difference in the overall distribution of
the two classes. Models in which most of the S0 galaxies are 
produced by major mergers involving starbursts in the recent past are
inconsistent with these observations since they predict that the merger 
remnant will have a higher blue luminosity up to several Gyrs
after the burst event and have an increased $\sigma$ as well. 
In contrast, the data are compatible
with less-extreme truncation models in which the cluster
environment suppresses star formation
rate in normal spirals (e.g.\ \pcite{LTC80,PSDCB99}; Balogh et al.\ 1999).

\subsubsection{Line Index Analysis}
The distributions of E and S0 galaxies in A\,2218 also appear
very similar in the \mgb--$\sigma$ diagram (Fig.\,\ref{sigmgb}). 
A KS test finds no significant differences  ($p=0.61$), and the
bootstrap bisector fits are compatible (Table~D1). This remains true 
even if the sample is restricted to {\it HST} classifications.
We see a similar picture in
the \mgb--\hb\ diagram (Fig.\,\ref{mghb}). The galaxies with strong
disks in A\,2218 are again distributed similarly to those
galaxies without disks (KS: $p=0.62$). 

Our analysis of the individual galaxies in A\,2218 supports the results
for the stellar populations in the composite, luminous elliptical and
S0 populations in clusters at $z=0.31$ from \scite{JSC00}.  In both
cases there appears to be little evidence for differences between the
stellar populations of the two samples.  This points  to a common
formation epoch for the bulk of the stars in most of the early--type
galaxies in A\,2218. Naively this appears  to be at odds with the strong
morphological evolution of cluster galaxies reported by
\scite{DOCSE97}. But the two findings can be reconciled with each other
if the suggested transformation from spirals to lenticulars does not
involve significant new star formation. 
The line indices are clearly sensitive to both the
strength of any past star formation event 
and the time which has elapsed since the last star formation.  Any model for the formation of S0 galaxies which
predicts a low enough ratio of these two parameters is viable \cite{PSDCB99,KS00}.

\subsubsection{Fundamental Plane}
The {\it HST} sample of twenty A\,2218 galaxies is split morphologically nearly 
evenly between elliptical and early-type disk galaxies (see Section\,\ref{obs}).  
These two subsamples are equally distributed across the 
surface of the plane so that the edge-on projection can be used to 
reliably compare their stellar populations. Both subsamples have lower
scatters than the combined sample: 0.077 in $\log r_e$ for the 9 Es and 0.091 for 
the 11 S0s, compared to 0.108 for all galaxies.
The lenticulars 
lie predominantly below the E galaxies with an average offset of 0.11$\pm$0.05
for almost identical slopes.
However, this need not reflect an evolutionary trend in the stellar
populations of E vs S0 galaxies. \scite{SBD93}
have reported a similar offset in galaxies in local
clusters. The offset may simply be caused by the way $R_e$ and $\langle
I\rangle_e$ are interrelated, resulting in a slightly higher surface
brightness for the disk galaxies at a given certain velocity dispersion
and effective radius.

\subsubsection{Discussion}

At low redshift,  E and S0 galaxies have similar
stellar populations (e.g.\ Bender et al.\ 1992, 1993) with only the highest 
signal-to-noise ratio line index analyses begining to show a distinction in the
star formation histories of the two types.
For example, \scite{Kunt00} found that the faint S0 in the Fornax cluster
tended to have younger ages relative to the (typically more luminous)
elliptical galaxies in the cluster. However, these differences might reflect
the mass dependence of star formation history since the majority of the 
S0 galaxies in Kuntschner's sample were systematically fainter than the
Fornax ellipticals. The two brightest and most massive S0 galaxies 
were found to have similar star formation histories to the ellipticals.

At intermediate reshift we should expect the differences between E and
S0 galaxies to become more apparent, particularly if the
morphological mix of the cluster population evolves rapidly with
look-back time \cite{DOCSE97}. 
However, our data suggest that the  stellar populations in
ellipticals and lenticulars are very similar
on {\it average}, although individual galaxies
introduce large variations in galaxy properties of {\it both} E and S0
populations.  There is little systematic difference between the two
samples in the \hb\ line index diagram or around the  \mgb--$\sigma$
relation.  In terms of the line index analysis both E and S0
galaxies  exhibit a similarly large range of ages. In our sample
these galaxies span a similar range of velocity dispersion, so that the
importance of mass and morphology can be distinguished: velocity
dispersion (or mass) seems to play a much more prominent role in
determining the stellar population of a galaxy than morphology alone.

Although ellipticals and lenticulars are similarly distributed within
the Fundamental Plane, the smaller S0 galaxies with lower velocity
dispersions are slightly offset from the average edge-on FP and have
smaller M/L values than the bulk of the ellipticals. This may arise
from either a somewhat increased luminosity (although the line index analysis
seems to preclude this) or from differences in the dynamics
which would produce a different relationship between mass,
$\sigma$ and $R_e$ for ellipticals and S0s.

\subsection{Radial Dependence}
\label{raddep}

Several authors have reported variations in the properties of cluster galaxies
with their distance from the cluster core. \scite{ASHCYEMOR96},
\scite{DFKIFF98} and
\scite{PKSECZO01} have found a radial dependence of the colour-magnitude
relation. Clearly, such studies need to be carefully controlled in order to 
take into account the density--morphology relation that is well known in both 
local and distant clusters \cite{DOCSE97}. In order to explore any  
 dependence on cluster radius within our dataset, we subdivide our sample into two
radial bins (at $r=130''$), such that they contain equal numbers of galaxies.
The average projected radius of galaxies in the outer bin is $230''$ (805\,kpc)
compared with $70''$ (245\,kpc) for galaxies in the core region.

\subsubsection{Faber--Jackson relation}
The central and outer galaxies have differing mean velocity
dispersions ($\log\sigma=2.30$ and 2.17 respectively) and
hence the two dimensional KS 
test is not the appropriate statistical method to compare these subsamples.
Therefore, we restrict our comparison to the bootstrap bisector fits. 
The slope for the galaxies in the outskirts is shallower than the
one for the core galaxies ($\Delta$(slope)$=1.16\pm1.17$), but not
significantly. The difference in mean luminosity is negligible.

\subsubsection{Line index anlaysis}
We find that  the inner and outer subsamples are equally distributed around
the \mgb--$\sigma$ fit ($\Delta$(slope)$=0.35\pm0.77$).
The median values of the two datasets fall within the range of fits for
the joint sample in Fig.\,\ref{sigmgb} revealing again the universality
of the \mgb--$\sigma$ relation. 

The bulk of the core galaxies occupy a small region within the \mgb--\hb\ plane,
corresponding to a relative narrow range of model ages and metallicities,
with the remainder spread over a wider region of the plane.
The galaxies located in the outskirts of the cluster, on the 
other hand, show a larger spread with both high and low \hb\ values and lower
\mgb\ line strengths (Fig.\,\ref{mghb}). Due to the systematic difference
in velocity dispersion it is hard to directly compare the distribution
on a statistical basis.

\subsubsection{Discussion}

\scite{ASHCYEMOR96} have analysed the colours of galaxies in the A\,2390 
cluster. They found a systematic trend in the colour--magnitude relation for
galaxies to become bluer in the outer parts
of the cluster, outside $\approx1$\,Mpc (see also
\pcite{PKSECZO01}). They interpreted this as a gradient
in the age of the early-type galaxies in the cluster. Similarly,
\scite{DFKIFF98}, studied the colours of disky galaxies in the $z=0.33$
cluster Cl\,1358$+$62 and reported a tendency for disky galaxies to have
increasingly younger ages beyond a radius of $0.7h_{50}^{-1}$\,Mpc. 

However, the change in the colours of galaxies with radial distance
might well be driven by the morphology-density relation: at larger
radius a greater fraction of the galaxy population will be late-type
spirals, and it may be this (rather than a bluing of the colours of
galaxies of a particular morphology) that drives the radial dependence
seen by \scite{DFKIFF98}.  At some level, the difference between a
disky, but bulge-strong galaxy (e.g.\ Sa) and a ``pure'' lenticular
is unimportant. For instance, if we are testing the formation and
evolution of the stellar population. However, the morphological
information can be used to tie together galaxies in a likely
evolutionary sequence, for example linking spiral galaxies in the
outer parts of the cluster to the formation of S0 galaxies.

In our study we sample galaxies over a similar radius to the colour-based
work, although good morphological imaging is only available for the core 
region. Overall there appears to be little difference between the radial 
subsamples, with both outer and inner region having galaxies
with a wide range
of stellar ages in the \hb\ diagram, for example. Thus old 
galaxies seem to be distributed throughout the cluster and not
limited only to the centre. These results are inconsistent with a simple
model in which the cluster grows a series of onion shells, with the oldest
galaxies being confined to the most bound orbits, and the most recently 
accreted galaxies are confined to orbits which tend to keep them away from the 
core. However, numerical simlations have shown that this picture is naive
and that cluster populations become mixed quite effectively over a period of only a 
few dynamical times \cite{MLB99}, particularly if the cluster goes through a
significant merging event. Thus the infall model cannot
be ruled out by this study due to the limited range of radii that we have 
covered. In order to make a definitive test of this model, spectra will
need to be obtained for galaxies out to 2--3 virial radii.

\section{Summary and conclusions}

We have found:
\begin{itemize}
\item The FJ and Mg-$\sigma$ relations have similar slopes to those
seen in Coma. The offset 
between the relations is small but is consistent with the difference 
in look-back time between the clusters if
the light of the galaxies are dominated by stars formed at $z_f>2$.

\item  The age-sensitive \hb\ index provides an alternative means to 
compare the stellar populations of galaxies in A\,2218 with local systems.
We find larger differences than for the FJ and Mg-$\sigma$ relations, implying
contamination by more recent star formation in some galaxies.

\item The \hb\ index also shows a large variation
between galaxies, with both E and S0 galaxies spanning a wide range in line strengths. The youngest galaxies are usually systems with lower
velocity dispersion. 

\item The distribution of E and S0 galaxies in age and metallicity as seen in
the \mgb--\hb\ diagram is quite similar. Thus, S0 galaxies are not restricted
to young ages and/or high metallicities. This suggests that higher
metallicities and younger ages are not necessarily conspiring to produce the small scatter observed in color-magnitude and FP relations
of early-type galaxies as was suggested by \scite{T97} and others.

\item The M/L ratios derived from the Fundamental Plane have a somewhat
steeper slope in A\,2218 than the Coma cluster, although
the effect is weaker than that seen in a
composite of intermediate redshift clusters by JFHD.
The increase in the slope of the relation is primarily driven by
the disk-type galaxies in A\,2218, a morphological distinction that
is not apparent in the line index diagrams.

\item Galaxies in both the central and outer samples span a wide range in ages, showing that
the stellar populations are not well correlated with radius.  We
will revisit this point in our next paper. 
\end{itemize}

These results agree well with the analysis of deep optical and
infrared photometry of this cluster presented in \scite{S00}.
Both studies find that the overall evolution of the population
is consistent with passive stellar evolution, but that faint galaxies show 
substantially greater diversity in their star formation histories than their 
bright counterparts.

Our current approach of
looking at the stellar populations of intermediate redshift clusters in
detail is directly complimentary to the approach taken by
\scite{SEEB98}, \scite{KB00} and others, of linking together the galaxy
populations of clusters at different redshifts in an evolutionary
sequence. A consensus model is emerging from these studies in which galaxies are
continually accreted from the surrounding field, their star formation
is strongly suppressed by the cluster and their stellar populations
generally age following passive evolutionary tracks.

The diversity of galaxy ages that we have found in this paper arises 
naturally in such a model. However, the model needs to be developed
to account for two features seen in our data. Firstly, the younger
galaxies are not uniquely disk (S0) systems, suggesting that the cluster
environment must disrupt the disk of the infalling galaxies as well as 
suppressing star formation. Secondly, old stellar populations dominate
in brighter galaxies, suggesting that most of the stars in bright 
galaxies were already in place at $z\gg 1$ (although not necessarily in a 
single system): the model must explain why the star formation histories of
bright and faint galaxies differ.  The key issue is whether the same
star formation histories (and their dependence on galaxy mass) hold in
groups and lower mass clusters. By investigating a much wider range
of environments, we will gain insight into
the decline of star formation in clusters, probing whether it is driven
by a cluster-specific mechanism (such as ram-pressure stripping, see e.g.
\pcite{A99,Q00}) or due to the decline in the gas reservoir available
to galaxies in the Universe as a whole.

We have carried out a similar study 
to this paper in the cluster Abell\,2390 ($z=0.23$). Since A\,2218 is at a
similar redshift and both clusters have high X-ray luminosities, we will
intercompare these two clusters in a forthcoming paper. For this purpose, we 
have also expanded our fiducial cosmological model and will present 
combined analyses
of line strength and mass-to-light ratio measurements which attempt to be
more independent of such models. Moreover, to  further increase the number of 
cluster galaxies with accurate morphological and structural parameters we will
utilize the recently completed wide-field mosaic taken by {\it HST}/WFPC2 of A\,2218.
Combining this with our panoramic spectroscopy will yield one of the largest
samples for studies of the evolution of the Fundamental Plane. This will allow us
to also explore how the large scatter seen in \hb\ line strengths for galaxies
with low velocity dispersions propagates into the Fundamental Plane.

\section*{Acknowledgments}

We thank Warrick Couch for kindly providing his visual classifications
for galaxies in A\,2218.  We acknowledge the anonymous referee for
her/his constructive review of our paper.  BLZ and DL acknowledge
support from PPARC, RGB from Durham University, IRS from the Royal
Society and RLD from the Leverhulme Trust.  This paper is based on
observations with the NASA/ESA Hubble Space Telescope which is operated
by the Space Telescope Science Institute under NASA contract
NAS5--26555, the William Herschel Telescope, which is operated by the
ING on behalf of PPARC and the Hale Telescope of Palomar Observatory,
which is owned and operated by Caltech.



\clearpage

\appendix
\onecolumn

\section[]{Ground--based photometric data}


%
%

%
\begin{figure}
\centerline{\psfig{figure=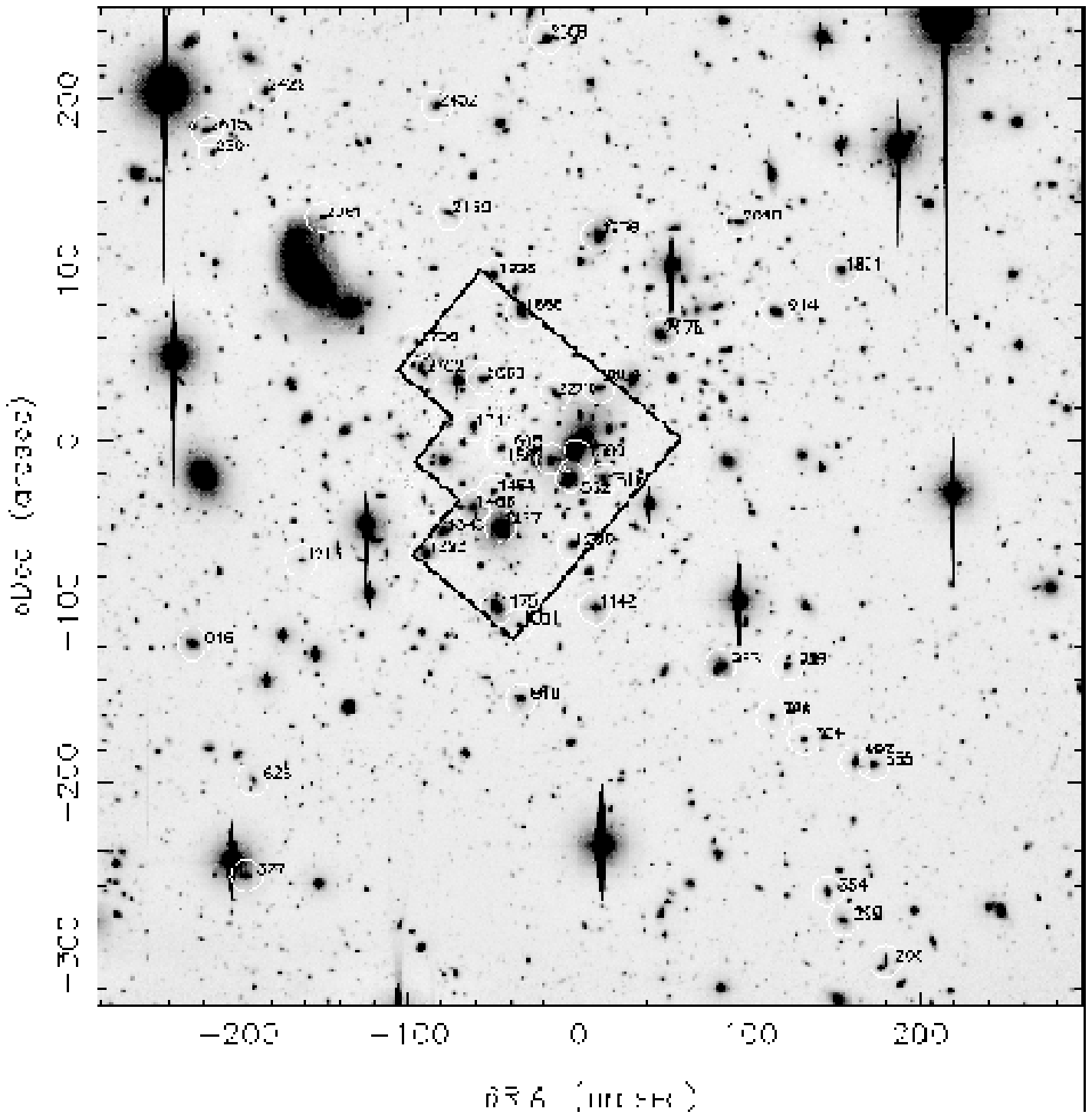,width=100mm,angle=0}} 
\caption{$I$-band image of the cluster A\,2218 taken with the  
5.1-m Hale telescope. The labels correspond to the galaxy numbers of
Table\,\ref{coomag}. North is up and East to the left. Coordinates are 
distance from the central galaxy in arcseconds. The {\it HST} field-of-view is
indicated as well.}
\label{finder}
\end{figure}

\newpage

\begin{figure*}
Place Fig. mosaic\_p200.jpg here.
\smallskip

Place Fig. mosaic\_p200b.jpg here.

\caption{$20''\times 20''$ sections from the Hale $I$-band images of
each of the spectroscopically confirmed cluster members in our sample.
These panels have North top and East left and the contours are in
0.5\,mag increments with the faintest contour being $\mu_I = 21.5$\,mag
arcsec$^{-2}$.  The labels give the galaxy ID as listed in
Table\,\ref{coomag} and where available the morphology from the {\it
HST} image.}
\label{mosaicp200}
\end{figure*}

\clearpage

\begin{table*}
\centering
\begin{minipage}{143mm}
\caption{}
\label{coomag}
\begin{tabular}{rccccccc}
ID & R.A. & Dec. & $I_{\mathrm{tot}}$ & $U_{\mathrm{ap}}$ & $B_{\mathrm{ap}}$
& $V_{\mathrm{ap}}$ & $I_{\mathrm{ap}}$\\
 & \multicolumn{2}{c}{(J2000)} &   &   &   &   &   \\
\hline
 208 & 16\,35\,19.89 & $+$66\,07\,41.5 & 17.224 & 21.431$\pm$0.053 & 21.031$\pm$0.004 & 19.611$\pm$0.003 & 17.832$\pm$0.002\\
 299 & 16\,35\,23.91 & $+$66\,08\,05.9 & 17.128 & 20.715$\pm$0.016 & 20.362$\pm$0.003 & 19.097$\pm$0.003 & 17.538$\pm$0.002\\
 354 & 16\,35\,25.42 & $+$66\,08\,22.3 & 17.751 & 21.706$\pm$0.012 & 21.208$\pm$0.004 & 19.789$\pm$0.003 & 18.151$\pm$0.002\\
 377 & 16\,36\,21.37 & $+$66\,08\,32.6 & 17.311 & 21.359$\pm$0.007 & 20.969$\pm$0.008 & 19.532$\pm$0.005 & 17.801$\pm$0.004\\
 626 & 16\,36\,20.64 & $+$66\,09\,28.3 & 17.509 & 21.426$\pm$0.009 & 21.082$\pm$0.004 & 19.541$\pm$0.003 & 17.914$\pm$0.002\\
 665 & 16\,35\,20.86 & $+$66\,09\,36.0 & 17.131 & 20.749$\pm$0.005 & 20.514$\pm$0.003 & 19.193$\pm$0.003 & 17.611$\pm$0.002\\
 697 & 16\,35\,22.66 & $+$66\,09\,37.7 & 17.142 & 21.079$\pm$0.007 & 20.725$\pm$0.004 & 19.230$\pm$0.003 & 17.571$\pm$0.002\\
 704 & 16\,35\,27.58 & $+$66\,09\,50.5 & 18.320 & 21.679$\pm$0.008 & 21.512$\pm$0.004 & 20.162$\pm$0.003 & 18.571$\pm$0.002\\
 786 & 16\,35\,30.67 & $+$66\,10\,04.5 & 17.950 & 21.618$\pm$0.007 & 21.295$\pm$0.004 & 19.795$\pm$0.002 & 18.154$\pm$0.001\\
 849 & 16\,35\,54.85 & $+$66\,10\,14.5 & 17.203 & 21.230$\pm$0.007 & 20.878$\pm$0.004 & 19.326$\pm$0.003 & 17.703$\pm$0.002\\
 926 & 16\,35\,29.09 & $+$66\,10\,33.8 & 17.900 & 21.912$\pm$0.008 & 21.510$\pm$0.005 & 19.997$\pm$0.004 & 18.374$\pm$0.002\\
 983 & 16\,35\,35.53 & $+$66\,10\,34.0 & 16.210 & 20.809$\pm$0.006 & 20.416$\pm$0.004 & 18.826$\pm$0.004 & 17.187$\pm$0.003\\
1046 & 16\,36\,26.42 & $+$66\,10\,47.3 & 16.824 & 20.961$\pm$0.006 & 20.642$\pm$0.003 & 19.086$\pm$0.003 & 17.472$\pm$0.002\\
1051 & 16\,35\,54.96 & $+$66\,10\,58.6 & 18.235 & 22.207$\pm$0.013 & 21.796$\pm$0.006 & 20.238$\pm$0.004 & 18.601$\pm$0.003\\
1142 & 16\,35\,47.59 & $+$66\,11\,07.8 & 16.618 & 20.634$\pm$0.005 & 20.298$\pm$0.003 & 18.773$\pm$0.003 & 17.206$\pm$0.002\\
1175 & 16\,35\,57.14 & $+$66\,11\,08.1 & 16.132 & 20.633$\pm$0.007 & 20.270$\pm$0.005 & 18.663$\pm$0.005 & 17.079$\pm$0.004\\
1213 & 16\,36\,15.88 & $+$66\,11\,35.9 & 17.785 & 21.452$\pm$0.007 & 21.144$\pm$0.004 & 19.608$\pm$0.002 & 18.035$\pm$0.001\\
1256 & 16\,35\,49.79 & $+$66\,11\,44.5 & 17.683 & 21.827$\pm$0.011 & 21.489$\pm$0.006 & 19.982$\pm$0.005 & 18.409$\pm$0.003\\
1293 & 16\,36\,03.95 & $+$66\,11\,40.0 & 16.711 & 21.038$\pm$0.007 & 20.642$\pm$0.004 & 18.972$\pm$0.003 & 17.333$\pm$0.002\\
1343 & 16\,36\,02.25 & $+$66\,11\,52.5 & 16.775 & 20.974$\pm$0.006 & 20.571$\pm$0.004 & 18.914$\pm$0.003 & 17.278$\pm$0.002\\
1437 & 16\,35\,56.74 & $+$66\,11\,55.2 & 15.403 & 21.200$\pm$0.011 & 20.776$\pm$0.009 & 19.097$\pm$0.009 & 17.461$\pm$0.008\\
1454 & 16\,35\,57.35 & $+$66\,12\,15.3 & 17.768 & 21.515$\pm$0.007 & 21.183$\pm$0.004 & 19.605$\pm$0.003 & 17.988$\pm$0.002\\
1466 & 16\,35\,59.34 & $+$66\,12\,06.3 & 16.415 & 21.237$\pm$0.008 & 20.826$\pm$0.005 & 19.224$\pm$0.004 & 17.598$\pm$0.003\\
1516 & 16\,35\,46.76 & $+$66\,12\,22.5 & 17.454 & 21.452$\pm$0.008 & 21.131$\pm$0.006 & 19.597$\pm$0.005 & 18.007$\pm$0.004\\
1552 & 16\,35\,49.92 & $+$66\,12\,23.4 & 16.514 & 20.995$\pm$0.008 & 20.578$\pm$0.006 & 18.955$\pm$0.005 & 17.312$\pm$0.005\\
1580 & 16\,35\,49.38 & $+$66\,12\,35.9 & 17.097 & 21.081$\pm$0.012 & 20.751$\pm$0.010 & 19.206$\pm$0.010 & 17.640$\pm$0.009\\
1605 & 16\,35\,56.65 & $+$66\,12\,40.9 & 17.992 & 21.406$\pm$0.007 & 21.154$\pm$0.004 & 19.778$\pm$0.003 & 18.315$\pm$0.002\\
1662 & 16\,35\,51.78 & $+$66\,12\,34.0 & 16.634 & 20.732$\pm$0.006 & 20.343$\pm$0.004 & 18.762$\pm$0.004 & 17.163$\pm$0.003\\
1711 & 16\,35\,59.31 & $+$66\,12\,53.2 & 17.293 & 21.269$\pm$0.007 & 20.919$\pm$0.004 & 19.357$\pm$0.003 & 17.760$\pm$0.002\\
1831 & 16\,35\,23.68 & $+$66\,14\,23.1 & 17.235 & 21.223$\pm$0.006 & 20.862$\pm$0.003 & 19.331$\pm$0.002 & 17.684$\pm$0.002\\
1888 & 16\,35\,54.60 & $+$66\,14\,00.2 & 16.561 & 20.882$\pm$0.007 & 20.585$\pm$0.006 & 19.044$\pm$0.005 & 17.488$\pm$0.004\\
1914 & 16\,35\,30.09 & $+$66\,13\,59.6 & 17.044 & 21.097$\pm$0.007 & 20.807$\pm$0.004 & 19.329$\pm$0.004 & 17.699$\pm$0.003\\
1928 & 16\,35\,57.21 & $+$66\,14\,21.4 & 17.337 & 21.221$\pm$0.008 & 20.910$\pm$0.004 & 19.336$\pm$0.003 & 17.709$\pm$0.002\\
1976 & 16\,35\,47.23 & $+$66\,14\,44.4 & 16.089 & 20.454$\pm$0.005 & 20.105$\pm$0.003 & 18.517$\pm$0.003 & 16.930$\pm$0.002\\
2061 & 16\,36\,13.77 & $+$66\,14\,54.7 & 18.165 & 21.945$\pm$0.014 & 21.688$\pm$0.009 & 20.258$\pm$0.009 & 18.664$\pm$0.004\\
2076 & 16\,35\,41.15 & $+$66\,13\,46.8 & 16.274 & 20.694$\pm$0.006 & 20.353$\pm$0.004 & 18.740$\pm$0.004 & 17.123$\pm$0.003\\
2090 & 16\,35\,33.51 & $+$66\,14\,51.6 & 17.462 & 21.232$\pm$0.006 & 20.949$\pm$0.004 & 19.480$\pm$0.003 & 17.898$\pm$0.002\\
2139 & 16\,36\,01.55 & $+$66\,14\,57.4 & 17.802 & 21.467$\pm$0.008 & 21.268$\pm$0.004 & 19.860$\pm$0.003 & 18.257$\pm$0.002\\
2270 & 16\,35\,51.39 & $+$66\,13\,11.9 & 17.428 & 21.512$\pm$0.009 & 21.107$\pm$0.004 & 19.531$\pm$0.003 & 17.925$\pm$0.002\\
2304 & 16\,36\,24.41 & $+$66\,15\,32.9 & 18.331 & 21.936$\pm$0.009 & 21.550$\pm$0.004 & 20.138$\pm$0.003 & 18.567$\pm$0.002\\
2452 & 16\,36\,02.77 & $+$66\,16\,00.1 & 17.387 & 21.318$\pm$0.008 & 21.083$\pm$0.004 & 19.604$\pm$0.003 & 17.936$\pm$0.002\\
2508 & 16\,35\,52.09 & $+$66\,16\,39.0 & 17.465 & 21.338$\pm$0.010 & 21.060$\pm$0.006 & 19.565$\pm$0.006 & 17.939$\pm$0.004\\
2604 & 16\,35\,47.16 & $+$66\,13\,15.8 & 17.106 & 21.138$\pm$0.006 & 20.808$\pm$0.004 & 19.217$\pm$0.003 & 17.605$\pm$0.002\\
2615 & 16\,36\,24.96 & $+$66\,15\,45.8 & 17.460 & 21.404$\pm$0.008 & 21.053$\pm$0.004 & 19.598$\pm$0.003 & 18.001$\pm$0.002\\
2660 & 16\,35\,58.27 & $+$66\,13\,21.2 & 17.895 & 21.824$\pm$0.010 & 21.549$\pm$0.004 & 20.000$\pm$0.003 & 18.402$\pm$0.002\\
2702 & 16\,36\,04.16 & $+$66\,13\,25.0 & 16.923 & 20.863$\pm$0.006 & 20.516$\pm$0.004 & 18.882$\pm$0.003 & 17.285$\pm$0.002\\
2738 & 16\,36\,04.76 & $+$66\,13\,42.2 & 18.271 & 21.939$\pm$0.010 & 21.663$\pm$0.005 & 20.107$\pm$0.003 & 18.485$\pm$0.002\\
\end{tabular}

\medskip
The galaxy ID can be found in the finding chart (Fig.\,\ref{finder}), 
$I_{\mathrm{tot}}$ 
is the total magnitude derived from the Hale images 
(SExtractor's {\sc best\_mag}), $U_{\mathrm{ap}}$, $B_{\mathrm{ap}}$, 
$V_{\mathrm{ap}}$, 
and $I_{\mathrm{ap}}$ are magnitudes within a circular aperture of
4.0\arcsec\ diameter measured in the seeing-convolved images.
None of the given magnitudes are corrected for extinction.
\end{minipage}
\end{table*}

\clearpage

\section{Kinematic and structural parameters for the {\it HST} subsample}

\phantom{blablabla}

\begin{figure*}
\centerline{\psfig{figure=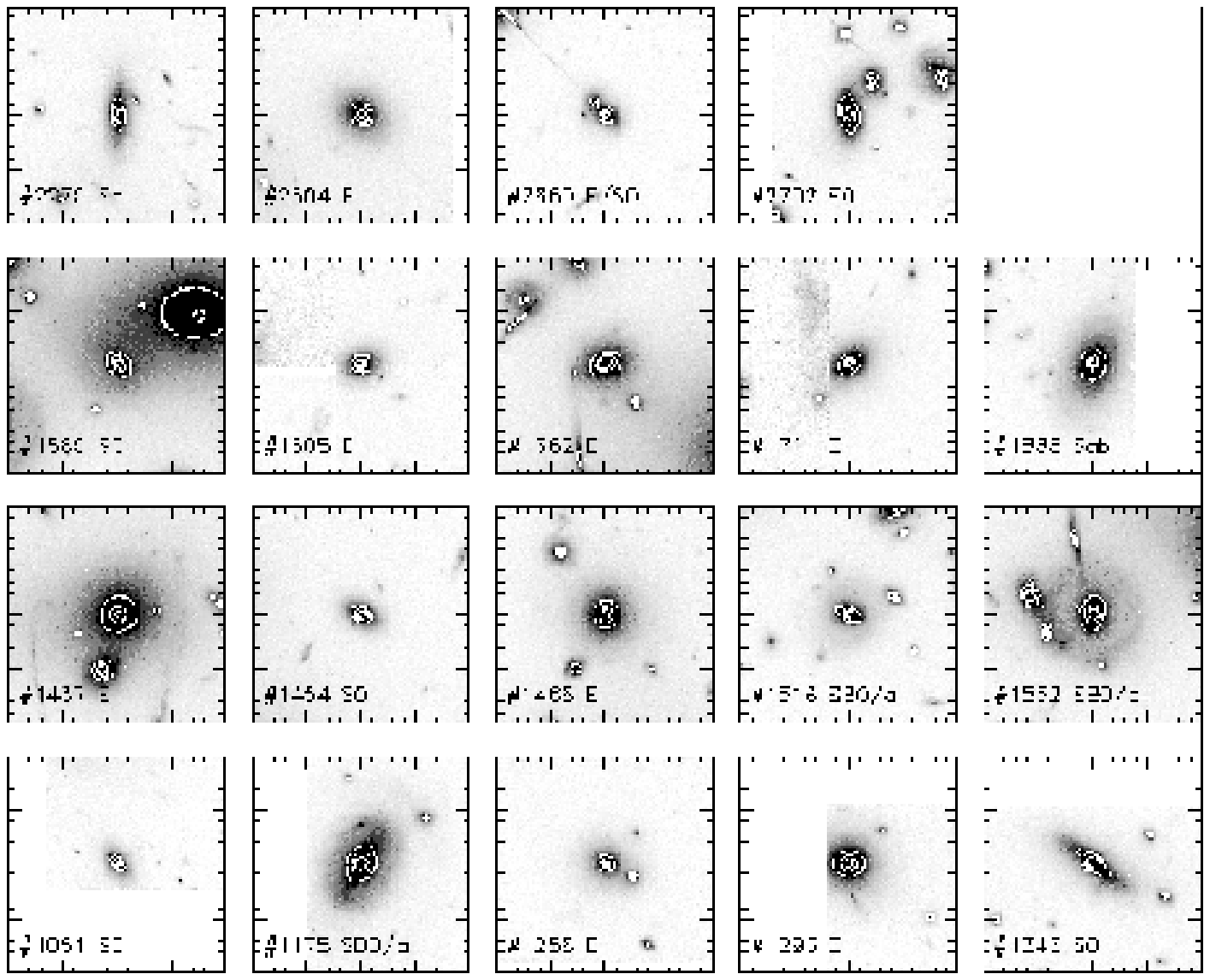,width=100mm,angle=0}}
\caption{Images of the 19 galaxies in our spectroscopic sample which fall 
within the \hst/WFPC2 field of A\,2218. Each panel is $20''$ square and is
labelled with the galaxy ID and visual morphology from Table\,\ref{hstpar}.}
\label{mosaichst}
\end{figure*}
\begin{table*}
\centering
\begin{minipage}{122mm}
\caption{}
\label{hstpar}
\begin{tabular}{rcccrcrccl}
ID & $\log(\sigma_a)$ & $R_{702}$ & $\log\langle I_{e}\rangle$
& $r_e$ & $\log(R_e)$ & $r_{e,b}$ & $h$ & $d/b$ & type \\
\hline
   0 & 2.50 & 14.81 &  1.43 &  18.347 & 1.85 & 30.00 &  2.75 & 0.27 & cD\\
1051 & 2.04 & 18.83 &  2.85 &   0.562 & 0.33 &  0.56 &  ...  & 0.00 & S0\\
1175 & 2.28 & 16.71 &  2.60 &   1.976 & 0.88 &  1.98 &  ...  & 0.00 & SB0/a\\
1256 & 2.18 & 18.29 &  2.26 &   1.421 & 0.74 &  1.12 &  1.75 & 0.22 & E\\
1293 & 2.38 & 17.42 &  2.92 &   0.997 & 0.58 &  0.75 &  0.82 & 0.49 & E\\
1343 & 2.38 & 17.26 &  3.01 &   0.968 & 0.57 &  1.70 &  0.20 & 0.33 & S0\\
1437 & 2.40 & 15.68 &  1.69 &   9.214 & 1.55 &  6.65 &  8.22 & 0.48 & E\\
1454 & 2.34 & 18.03 &  2.90 &   0.771 & 0.47 &  0.48 &  3.23 & 0.27 & S0\\
1466 & 2.30 & 17.23 &  2.37 &   2.055 & 0.90 &  2.05 &  ...  & 0.00 & E\\
1516 & 2.07 & 18.15 &  2.88 &   0.745 & 0.46 &  0.75 &  ...  & 0.00 & SB0/a\\
1552 & 2.23 & 16.89 &  2.45 &   2.198 & 0.93 &  2.56 &  0.10 & 0.08 & SB0/a\\
1580 & 2.21 & 17.91 &  2.98 &   0.732 & 0.45 &  0.73 &  ...  & 0.00 & S0\\
1605 & 2.13 & 18.51 &  2.70 &   0.761 & 0.47 &  0.59 &  0.63 & 0.43 & E\\
1662 & 2.56 & 17.08 &  2.90 &   1.186 & 0.66 &  1.60 &  0.11 & 0.16 & E\\
1711 & 2.31 & 17.95 &  2.90 &   0.794 & 0.48 &  0.59 &  0.95 & 0.29 & E\\
1888 & 2.19 & 17.44 &  2.62 &   1.380 & 0.72 &  0.38 &  1.15 & 2.02 & Sab\\
2270 & 2.23 & 17.96 &  2.63 &   1.082 & 0.62 &  1.08 &  ...  & 0.00 & Sa\\
2604 & 2.27 & 17.67 &  2.88 &   0.931 & 0.55 &  0.93 &  ...  & 0.00 & E\\
2660 & 2.08 & 18.68 &  2.48 &   0.926 & 0.55 &  0.93 &  ...  & 0.00 & E/S0\\
2702 & 2.35 & 17.49 &  3.10 &   0.782 & 0.48 &  0.78 &  0.47 & 0.21 & S0\\

\end{tabular}

\medskip
The galaxy ID can be found in the finding chart (Fig.\,\ref{finder})
and coordinates are given in Table\,\ref{coomag}, $\log(\sigma_a)$ is
the logarithm of the aperture corrected velocity dispersion
(the value for the cD galaxy \#\,0 was taken from \pcite{JFHD99}),
$R_{702}$ is the extinction-corrected, Vega-based total magnitude in the WFPC2 F702W
filter, $\log\langle I_e\rangle$ is the mean surface brightness within
the effective radius $r_e$ (in arcsec) in the $I$ band according to 
Equation\,8, $\log(R_e)$ is the logarithm of the
effective radius (in kpc), $r_{e,b}$ is the effective radius of the
bulge, $h$ is the scale-length of the disk (both in arcsec) and $d/b$ is
the luminosity weighted disk-to-bulge ratio (see Section\,\ref{secfp}
for the derivation). The galaxy types were obtained visually by Prof. Couch.
Thumbnail images are shown in Fig.\,\ref{mosaichst}.
\end{minipage}
\end{table*}

\clearpage

\section{Spectroscopic data}

\begin{table*}
\centering
\begin{minipage}{95mm}
\caption{}
\label{linpar1}
\begin{tabular}{rccccrc}
ID & $\sigma$ & $v$ & {H$\beta$} & 
{Mg$_b$} & S/N & disk\\
\hline
 208 & 181.1$\pm$5.6 & 46505$\pm$18 & 1.62$\pm$0.14 & 4.17$\pm$0.15 &    50 & d\\
 299 & 163.3$\pm$5.0 & 45708$\pm$15 & 2.64$\pm$0.11 & 2.72$\pm$0.12 &   133 &  \\
 354 & 105.5$\pm$7.2 & 48949$\pm$28 & 2.49$\pm$0.15 & 3.30$\pm$0.16 &    65 &  \\
 377 & 247.0$\pm$4.9 & 48505$\pm$17 & 2.03$\pm$0.18 & 4.10$\pm$0.20 &   137 &  \\
 626 & 128.1$\pm$5.3 & 46573$\pm$19 & 1.97$\pm$0.21 & 3.98$\pm$0.24 &    84 &  \\
 665 & 126.1$\pm$7.2 & 46577$\pm$22 & 2.09$\pm$0.13 & 2.54$\pm$0.14 &    99 &  \\
 697 & 189.5$\pm$8.6 & 49622$\pm$12 & 2.05$\pm$0.11 & 3.84$\pm$0.13 &    76 & d\\
 704 & 105.6$\pm$8.7 & 47276$\pm$17 & 2.49$\pm$0.24 & 3.27$\pm$0.26 &    70 &  \\
 786 & 137.1$\pm$4.9 & 50537$\pm$10 & 1.77$\pm$0.16 & 3.64$\pm$0.17 &    93 &  \\
 849 & 204.6$\pm$8.6 & 45484$\pm$18 & 1.37$\pm$0.11 & 4.34$\pm$0.13 &    82 &  \\
 926 & 132.6$\pm$4.8 & 49293$\pm$14 & 2.34$\pm$0.22 & 3.98$\pm$0.23 &    68 &  \\
 983 & 198.3$\pm$5.5 & 48917$\pm$17 & 1.47$\pm$0.09 & 4.40$\pm$0.10 &   132 &  \\
1046 & 247.1$\pm$4.8 & 46372$\pm$15 & 1.80$\pm$0.13 & 4.33$\pm$0.14 &    41 &  \\
1051 & 105.6$\pm$4.9 & 48272$\pm$23 & 1.72$\pm$0.24 & 4.23$\pm$0.26 &    53 & d\\
1142 & 194.2$\pm$5.3 & 45491$\pm$15 & 1.93$\pm$0.09 & 3.88$\pm$0.10 &   113 &  \\
1175 & 182.8$\pm$5.2 & 48256$\pm$14 & 2.03$\pm$0.08 & 3.94$\pm$0.09 &   140 & d\\
1213 & 150.1$\pm$4.6 & 48067$\pm$22 & 1.87$\pm$0.18 & 3.99$\pm$0.21 &    75 &  \\
1256 & 145.7$\pm$5.1 & 49638$\pm$ 1 & 1.91$\pm$0.22 & 3.97$\pm$0.23 &    72 &  \\
1293 & 229.1$\pm$4.8 & 47896$\pm$17 & 1.60$\pm$0.11 & 4.78$\pm$0.13 &   109 &  \\
1343 & 230.6$\pm$4.9 & 50027$\pm$15 & 1.87$\pm$0.11 & 4.67$\pm$0.12 &   110 & d\\
1437 & 237.3$\pm$4.5 & 48495$\pm$15 & 1.76$\pm$0.10 & 4.74$\pm$0.11 &   127 &  \\
1454 & 210.0$\pm$5.5 & 49624$\pm$ 9 & 1.77$\pm$0.19 & 4.40$\pm$0.21 &   103 & d\\
1466 & 189.9$\pm$5.1 & 49405$\pm$12 & 1.88$\pm$0.14 & 4.17$\pm$0.15 &   107 &  \\
1516 & 111.8$\pm$7.0 & 44919$\pm$13 & 2.15$\pm$0.15 & 3.40$\pm$0.16 &    78 & d\\
1552 & 161.4$\pm$5.4 & 48842$\pm$19 & 2.05$\pm$0.11 & 3.78$\pm$0.13 &    43 & d\\
1580 & 155.5$\pm$4.7 & 47850$\pm$14 & 2.05$\pm$0.13 & 3.81$\pm$0.14 &   122 & d\\
1605 & 128.3$\pm$4.7 & 47050$\pm$17 & 2.67$\pm$0.18 & 3.57$\pm$0.19 &    86 &  \\
1662 & 348.5$\pm$5.6 & 44948$\pm$28 & 1.80$\pm$0.09 & 5.35$\pm$0.11 &   150 &  \\
1711 & 194.5$\pm$5.4 & 47619$\pm$20 & 1.81$\pm$0.15 & 4.60$\pm$0.17 &    89 &  \\
1831 & 182.2$\pm$5.5 & 47833$\pm$14 & 1.84$\pm$0.15 & 4.40$\pm$0.17 &    80 &  \\
1888 & 147.1$\pm$5.3 & 47906$\pm$ 4 & 2.50$\pm$0.11 & 3.29$\pm$0.12 &    86 & d\\
1914 & 140.8$\pm$5.1 & 47556$\pm$15 & 1.65$\pm$0.16 & 4.20$\pm$0.18 &    87 &  \\
1928 & 207.0$\pm$4.6 & 48684$\pm$14 & 2.00$\pm$0.12 & 3.88$\pm$0.13 &   137 &  \\
1976 & 266.2$\pm$5.0 & 46339$\pm$17 & 1.56$\pm$0.08 & 4.48$\pm$0.09 &   145 &  \\
2061 & 131.5$\pm$8.2 & 49002$\pm$ 8 & 1.65$\pm$0.24 & 3.68$\pm$0.26 &    73 &  \\
2076 & 209.1$\pm$5.2 & 49399$\pm$15 & 1.63$\pm$0.09 & 4.16$\pm$0.10 &   138 &  \\
2090 & 134.4$\pm$5.3 & 49211$\pm$15 & 2.13$\pm$0.19 & 3.81$\pm$0.21 &    86 & d\\
2139 & 114.5$\pm$6.5 & 48905$\pm$ 2 & 1.96$\pm$0.18 & 3.14$\pm$0.19 &    59 & d\\
2270 & 160.6$\pm$5.2 & 46078$\pm$12 & 1.39$\pm$0.20 & 4.26$\pm$0.22 &    79 & d\\
2304 & 101.6$\pm$4.7 & 46568$\pm$17 & 3.53$\pm$0.23 & 3.48$\pm$0.25 &    60 &  \\
2429 & 163.3$\pm$4.9 & 46235$\pm$12 & 2.12$\pm$0.20 & 3.80$\pm$0.22 &    46 &  \\
2452 & 127.3$\pm$4.8 & 46286$\pm$16 & 1.66$\pm$0.17 & 3.92$\pm$0.19 &    68 & d\\
2508 & 128.0$\pm$5.3 & 47114$\pm$17 & 1.47$\pm$0.14 & 3.87$\pm$0.15 &   102 &  \\
2604 & 179.3$\pm$9.9 & 49390$\pm$17 & 1.97$\pm$0.12 & 4.22$\pm$0.13 &    86 &  \\
2615 & 155.0$\pm$5.2 & 49558$\pm$12 & 2.11$\pm$0.16 & 3.32$\pm$0.17 &    83 &  \\
2660 & 114.8$\pm$4.5 & 48561$\pm$15 & 1.42$\pm$0.23 & 4.18$\pm$0.25 &    63 & d\\
2702 & 212.3$\pm$5.5 & 47747$\pm$14 & 1.89$\pm$0.09 & 4.39$\pm$0.10 &   156 & d\\
2738 & 173.7$\pm$5.5 & 47723$\pm$14 & 1.86$\pm$0.21 & 4.18$\pm$0.23 &    76 &  \\
\end{tabular}

\medskip
The galaxy ID can be found in the finding chart (Fig.\,\ref{finder}) and
coordinates are given in Table\,\ref{coomag}, $\sigma$ is 
the velocity dispersion measured in km\,s$^{-1}$ (not aperture corrected) and 
its error, $v$ the radial velocity and its error. The line indices are measured in the 
flux-calibrated spectra and given in \AA. The signal-to-noise ratio is given 
in the last but one column,
while the last column indicates whether the galaxy has a strong disk (d).
\end{minipage}
\end{table*}

\clearpage

\begin{figure*}
\centerline{\psfig{figure=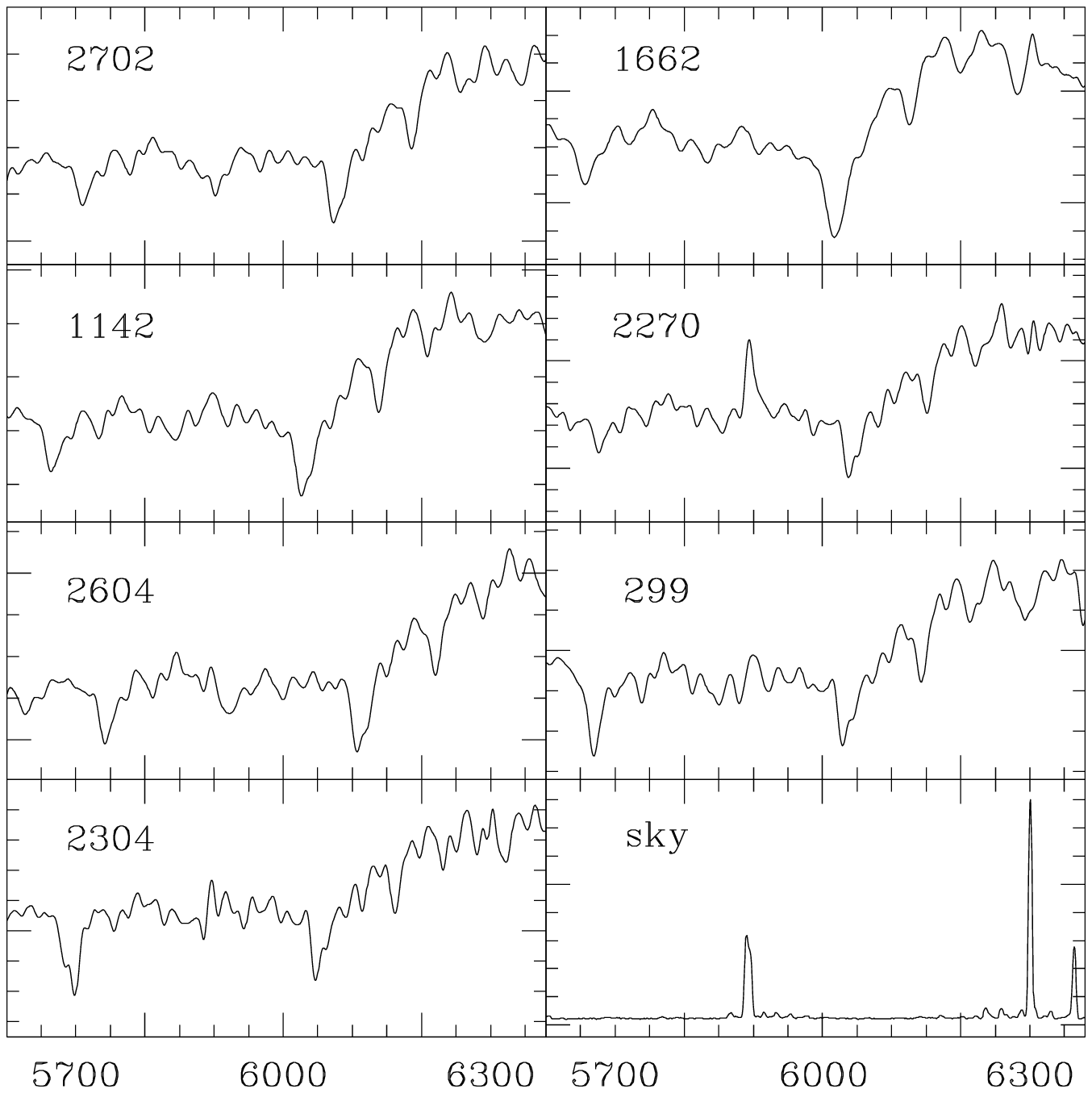}}
\caption{A\,2218 example spectra shown in the observed frame. Spectra were
  degraded to the Lick resolution for measurement of the indices (for 
identification, see Fig.\,\ref{compspec}). The left panels show 
spectra with decreasing S/N ($\sigma$) from top to bottom, whereas the right
panels show \lq peculiar\rq\ galaxies (1662: high $\sigma$, 2270: low \hb,
299: low \mgb) and the skylines.}
\label{spectrum}
\end{figure*}

\clearpage

\section{Bootstrap bisector fits}

\label{lastpage}

\begin{table*}
\centering
\begin{minipage}{132mm}
\caption{}
\label{fitpar}
\begin{tabular}{lcccccc}
Relation & Local & A\,2218: Full & E &        S0 &        In &       Out\\
\hline
(1) $\log\sigma$--\mgb & 3.80$\pm$0.35   &3.97$\pm$0.49   &4.18$\pm$0.67   &3.15/1.01  &3.61$\pm$0.57   &3.96$\pm$0.52\\
           & 4.40$\pm$0.06   &4.33$\pm$0.06   &4.28$\pm$0.08   &4.40$\pm$0.09   &4.39$\pm$0.09   &4.27$\pm$0.09\\
           &            &$-$0.07      &$-$0.12      &0          &$-$0.01      &$-$0.13\\
           & $$-$$4.10      &$-$4.55      &$-$5.08      &$-$2.66      &$-$3.70      &$-$4.59\\
           & 72         &48         &31         &16         &24         &24\\
                                                                         \\
(1a) $\log\sigma$--\mgb&            &3.54$\pm$0.31   &3.37$\pm$0.39   &           &           &\\
           &            &4.39$\pm$0.04   &4.37$\pm$0.05   &           &           &\\
           &            &$-$0.01      &$-$0.03      &           &           &\\
           &            &$-$3.53      &$-$3.18      &           &           &\\
           &            &46         &29         &           &           &\\
                                                                         \\
(2) $\log\sigma$--$r_{\rm abs}$& $-$6.82$\pm$0.89  &$-$5.10$\pm$0.69  &$-$5.44$\pm$0.35  &$-$4.32$\pm$0.99  &$-$5.46$\pm$0.90  &$-$4.30$\pm$0.75\\
           & $-$21.49$\pm$0.06 &$-$21.80$\pm$0.07 &$-$21.78$\pm$0.06 &$-$21.95$\pm$0.13 &$-$21.81$\pm$0.14 &$-$21.77$\pm$0.08\\
           &            &$-$0.31      &$-$0.29      &$-$0.45      &$-$0.32      &$-$0.28\\
           & $-$6.23      &$-$10.38     &$-$9.59      &$-$12.28     &$-$9.60      &$-$12.14\\
           & 75         &48         &31         &16         &24         &24\\
                                                                         \\
(3) $\log(R_e)$--FP& 1.05$\pm$0.05   &1.13$\pm$0.08   &1.08$\pm$0.06   &1.05$\pm$0.18   &           &\\
           & 0.58$\pm$0.01   &0.52$\pm$0.02   &0.60$\pm$0.03   &0.49$\pm$0.04   &           &\\
           &            &$-$0.06      &0.02       &$-$0.11      &           &\\
           & 0.03       &$-$0.07      &0.03       &$-$0.06      &           &\\
           & 88         &20         &9          &11         &           &\\
                                                                         \\
(4) $\log\sigma$--$\log(M/L)$ & 1.27$\pm$0.13   & 1.61$\pm$0.17  &1.52$\pm$0.42   &1.18$\pm$0.31   &           &\\
           & 0.60$\pm$0.01   & 0.52$\pm$0.02  &0.58$\pm$0.05   &0.47$\pm$0.04   &           &\\
           &            & $-$0.08     &$-$0.02      &$-$0.13      &           &\\
           & $-$2.29      & $-$3.16     &$-$2.88      &$-$2.23      &           &\\
           & 102        & 20        &9          &11         &           &\\
\end{tabular}

\medskip
1. row: slope$\pm1\sigma$\\
2. row: $y(\langle x\rangle)$ i.e. ordinate value at median value of abscissa\\
3. row: $\Delta$(A\,2218--local)\\ 
4. row: zeropoint\\
5. row: number of galaxies\\
Local sample: (1) SMAC (full sample),
(2) J{\o}ergensen reduced to value range of A\,2218: 
$\log(\sigma)>2.0$, $r_{\rm abs}<-20.7$,
(3) J{\o}ergensen reduced to value range of A\,2218: 
$\log(R_e)>0.26$,FP$>0.20$,
(4) J{\o}ergensen reduced to value range of A\,2218: 
$\log(\sigma)>2.0$\\
Distant sample: A\,2218 full sample; E: only ellipticals; S0: only lenticulars;
In: only core sample; Out: only outer region sample;
(1a) full sample without \#299 and \#665
\end{minipage}
\end{table*}
%


\end{document}